%% file: main.tex
\documentclass[12pt]{article}

\usepackage{graphicx}
\usepackage{bbm}
\usepackage{amsmath}
\usepackage{float}
\usepackage{xcolor}
\usepackage{subcaption}
\usepackage[margin=1in]{geometry}
\usepackage[citestyle=authoryear,natbib=true]{biblatex}
\usepackage{setspace}
\usepackage{hyperref}
\usepackage{tabularx}
\usepackage[flushleft]{threeparttable}
\usepackage{soul}

\newcommand{\PreserveBackslash}[1]{\let\temp=\\#1\let\\=\temp}
\newcolumntype{C}[1]{>{\PreserveBackslash\centering}p{#1}}
\newcolumntype{R}[1]{>{\PreserveBackslash\raggedleft}p{#1}}
\newcolumntype{L}[1]{>{\PreserveBackslash\raggedright}p{#1}}

\title{The OxyContin Reformulation Revisited: New Evidence From Improved Definitions of Markets and Substitutes}
\author{Shiyu Zhang\thanks{Zhang: California Institute of Technology. Corresponding author, email szzhang@caltech.edu. Guth: California Institute of Technology}
\and Daniel Guth}
\date{\today}

\begin{document}
\onehalfspacing

\begin{titlepage}
\maketitle
\begin{abstract}

    \noindent The opioid epidemic began with prescription pain relievers. In 2010 Purdue Pharma reformulated OxyContin to make it more difficult to abuse. OxyContin misuse fell dramatically, and concurrently heroin deaths began to rise. Previous research overlooked generic oxycodone and argued that the reformulation induced OxyContin users to switch directly to heroin. Using a novel and fine-grained source of all oxycodone sales from 2006-2014, we show that the reformulation led users to substitute from OxyContin to generic oxycodone, and the reformulation had no overall impact on opioid or heroin mortality. In fact, generic oxycodone, instead of OxyContin, was the driving factor in the transition to heroin. Finally, we show that by omitting generic oxycodone we recover the results of the literature. These findings highlight the important role generic oxycodone played in the opioid epidemic and the limited effectiveness of a partial supply-side intervention. 

    \vspace{0in}
    \noindent\textbf{Keywords:} Opioids, Drug Overdoses, Heroin, Drug Distributors\\
    \vspace{0in}\\
    \noindent\textbf{JEL Codes:} I11, I12, I18 \\
    \bigskip
\end{abstract}
\end{titlepage}

\section{Introduction}
Since 1999, the opioid epidemic has claimed more than 415,000 American lives (CDC Wonder). What started with fewer than 6,000 opioid-related deaths in 1999 grew steadily every year until fatalities reached 47,573 deaths in 2017. Following a small decline in fatal drug overdoses in 2018, deaths continue to rise. Over the past two decades, millions of Americans have misused prescription opioids or progressed to more potent opioids, first heroin and later fentanyl. Many social scientists have tried to understand how this crisis has grown over two decades despite significant public health efforts to the contrary. 

Doctors and health economists have long argued that the drug most responsible for prescription opioid overdose deaths, and the key to understanding the transition from prescription opioids to heroin starting in 2010, was OxyContin. Previous research \parencite{van2009promotion}, court proceedings \parencite{meier2007guilty}, and books (\cite{meier2003pain},  \cite{macy2018dopesick}) has documented how Purdue Pharma's marketing campaign for OxyContin downplayed the risk of addiction starting in 1996. Since then, according to the National Survey on Drug Use and Health (NSDUH), millions of Americans have misused it previously.  A key question in this area is whether or not making prescription opioids, especially OxyContin, more difficult to abuse will reduce overdose deaths.

In this paper, we show that restricting access to OxyContin led many users to switch to generic oxycodone but had no impact on opioid or heroin mortality. Earlier analyses attributing opioid overdose deaths in the late 2000s and the subsequent rise in heroin deaths to OxyContin are incomplete because they omit generic oxycodone. Our analysis shows that the misuse of generic oxycodone was prevalent before the reformulation that restricted OxyContin access, and was even more so afterward. We also show that heroin overdose deaths increased in areas with high generic oxycodone exposure, not high OxyContin exposure, two years after the OxyContin reformulation. In addition, we show that omitting generic oxycodone in our regressions recovers the results of the literature. 

This analysis was not possible until one year ago when The Washington Post won a court order and published the complete Automation of Reports and Consolidated Orders System  \href{https://www.washingtonpost.com/investigations/little-known-generic-drug-companies-played-central-role-in-opioid-crisis-documents-reveal/2019/07/26/95e08b46-ac5c-11e9-a0c9-6d2d7818f3da_story.html}{(ARCOS)}. The ARCOS tracks the manufacturer, the distributor and the pharmacy of every pain pill sold in the United States. The newly released data allow us to analyze what happened to sales of generic oxycodone and OxyContin when OxyContin suddenly became more difficult to abuse. The previous literature focused on analyzing OxyContin because of Purdue's notorious role in the opioid crisis. However, the new data shows that the sales of OxyContin was only a small part of the sales of all prescription opioids: in terms of the number of pills, OxyContin was \href{https://www.washingtonpost.com/investigations/an-onslaught-of-pills-hundreds-of-thousands-of-deaths-who-is-accountable/2019/07/20/8d85e650-aafc-11e9-86dd-d7f0e60391e9_story.html}{3\%} of all oxycodone pills sold from 2006 to 2012; in terms of morphine milligram equivalents (MME), OxyContin has closer to 20\% market share over this period. The new transaction-level ARCOS data allows us to track the sales of generic oxycodone and fill in the narrative gaps of how the opioid crisis progressed in the United States.

Following \citet{alpert2018supply}, \citet{evans2019reformulation} and \citet{Cicero2015}, we treat the introduction of an abuse-deterrent formulation (ADF) of OxyContin as an exogenous shock that should only affect people who seek to bypass the extended-release mechanism for a more immediate high. We construct measures of exposure by combining ARCOS sales and the NSDUH data on drug misuse. The NSDUH is the best survey of people who use drugs at the state level, and by combining it with local sales we can capture variation in drug use within the state. We leverage this variation in OxyContin and generic oxycodone exposure to examine how the reformulation affected OxyContin sales, generic oxycodone sales, opioid mortality, and heroin mortality. Our first contribution is that we fix the omitted-variable problem by differentiating between OxyContin and generic oxycodone, and we show that this leads to different conclusions than what previous literature suggests. Our second contribution is disaggregating the data to metropolitan statistical area (MSA), which allows us to address endogeneity at the state level. 


To preview our results, we find strong evidence of substitution from OxyContin to generic oxycodone immediately after the reformulation. This substitution was larger in places that had more OxyContin misuse pre-reform, which is consistent with our hypothesis that users would switch between oxycodones rather than move on to heroin. Because this substitution should be concentrated among people misusing OxyContin, the results imply large changes in consumption at the individual level. Back-of-the-envelope calculation suggests 68\% of the decline in OxyContin sales was substituted to oxycodone in MSAs with high OxyContin misuse. The findings are consonant with surveys like \citet{havens2014impact}, \citet{coplan2013changes}, and \citet{cassidy2014changes} who all document substitution to generic oxycodone after the reformulation by people seeking to bypass the ADF. We also find suggestive evidence of substitution from generic oxycodone to OxyContin after the reformulation in places where generic oxycodone misuse was high, a channel that has been unexplored in previous research. 

Our event study approach also shows that generic oxycodone exposure is predictive of future heroin overdose deaths whereas OxyContin exposure is not. The results are not contingent on methodology or our construction of exposure measures. Crucially, if we run the same exact regressions at the state or MSA level and omit generic oxycodone, we recover the results of the literature where OxyContin misuse appears to be significantly predictive of future heroin overdose deaths. We find that every standard deviation increase in generic oxycodone exposure pre-reformulation is associated with a 40.8\% increase in heroin mortality in 2012 from the 2009 baseline level. As further evidence against the argument that there was immediate substitution from OxyContin to heroin after the reformulation, we note that in all of our regressions the increase in heroin deaths wasn't statistically significant until 2012. As suggested in \citet{o2017trends}, the rise in heroin deaths can be attributed in part to an increase in the supply of heroin as well as the introduction of fentanyl into heroin doses. 

Our findings highlight the pitfalls of omitting important substitutes to OxyContin in analyzing the prescription opioid crisis. Purdue Pharma has received well-deserved attention over the years for its role in igniting the crisis. The company has been involved in many lawsuits over the years, but perhaps the most damaging were lesser-known cases that involved losing its patent in 2004\footnote{\href{https://www.nytimes.com/2004/01/06/business/judge-says-maker-of-oxycontin-misled-officials-to-win-patents.html}{Federal ruling}, \href{https://www.drugtopics.com/view/generic-oxy-makers-too-must-offer-risk-management}{Risk management plan proposals for generic oxycodone}} which cleared the way for a rapid increase in generic oxycodone sales in the early 2000s. While Purdue Pharma was being sued and scrutinized, several manufacturers took the opportunity to fill in the gaps of OxyContin. By 2006, generic oxycodone outsold OxyContin by more than 3-to-1 after accounting for pill dosage differences. This paper sheds lights on the role generic oxycodone played and continues to play in the opioid crisis and helps policy makers update their picture of the opioid use disorder (OUD) landscape.

The paper also calls attention to the limited effectiveness of a partial supply-side intervention to curb OUD. Purdue Pharma was once a dominant player in the opioid market, but by the time of the reformulation, that dominance had vanished and it was only one of the many manufacturers whose drugs were actively misused by Americans. Purdue was the first company to include abuse-deterrent formulation (ADF) in their opioids, but it is not until recent years that other brands started adding anti-deterrent compounds to their 
products \parencite{pergolizzi2018abuse}. When substitutions to other abusable opioids are easy, cutting supplies of one kind is less effective.

The rest of the paper runs as follows. Section 2 gives more background on the opioid crisis and explains how previous research has characterized the OxyContin reformulation. In Section 3 we describe the new ARCOS sales database, the NSDUH misuse data, the NVSS mortality data, as well as our constructed misuse measure and descriptive statistics. Section 4 describes our empirical strategy for testing our hypotheses. Section 5 discusses our results and what it means for our understanding of the transition between illicit drugs, and Section 6 concludes.

\section{Background and Literature Review}

This section proceeds in chronological order. First, we provide a history of oxycodone and its most important formulation, OxyContin. We then describe the OxyContin reformulation in 2010 and what it meant for prescription opioid misuse, as well as how the previous literature analyzed the reformulation. Next, we present the nascent research on substitution between different opioids and how our contribution fits in this strain of work. We conclude with a summary of the literature on heroin mortality in the early 2010s and its link with the prescription opioid crisis. 

Oxycodone was first marketed in the United States as Percodan by DuPont Pharmaceuticals in 1950. It quickly found to be as addictive as morphine \parencite{bloomquist1963addiction}, and in 1965 California placed it on the triplicate prescription form \parencite{quinn1965percodan}.\footnote{Triplicate programs required pharmacists to send a copy to the government, and \cite{alpert2019origin} show that these had a persisting effect on reducing the number of opioid prescriptions.} Before the 1990s, doctors were hesitant to prescribe oxycodone to non-terminally ill patients due to its high abuse potential \parencite{deweerdt2019tracing}. The sales of oxycodone-based pain relievers did not take off until the mass marketing of OxyContin, Purdue's patented oxycodone-based painkiller. OxyContin was first approved by the FDA in 1995. The drug's innovation was an `extended-release' formula, which allowed the company to pack a higher concentration of oxycodone into each OxyContin pill and the patients to take the pills every 12 hours instead of 8 hours. OxyContin's original label, approved by the FDA, stated that the ``delayed absorption, as provided by OxyContin tablets, is believed to reduce the abuse liability of a drug." In 2001, the FDA changed OxyContin's label to include stronger warnings about the potential for abuse and Purdue agreed to implement a Risk Management Program to try and reduce OxyContin misuse.\footnote{\href{https://www.fda.gov/media/126835/download}{From the FDA Opioid Timeline.}}

OxyContin was one of the first opioids marketed specifically for non-cancer pain. In the early 1990s, pain started to enter the medical discussion as the `fifth vital sign' and something to be managed. As described in \citet{meier2003pain}, \citet{van2009promotion}, and elsewhere, Purdue's sales representatives pushed OxyContin and were told to downplay the risk of addiction. \citet{quinones2015dreamland} describes how Purdue cited a 1980 short letter published in the New England Journal of Medicine describing extremely low rates of opioid addiction among hospital patients undergoing hospital stays, but the company repeatedly implied this result extended to the general population or to individuals who left the hospital with take-home prescriptions of OxyContin. The short letter was uncritically or incorrectly cited 409 times as evidence that addiction was rare with long-term opioid therapy (\cite{leung20171980}). As a result of Purdue's aggressive marketing and downplaying of the drug's abuse potential, OxyContin was a huge financial success and effectively catalyzed the prescription opioid crisis. 


In May 2007, Purdue signed a guilty plea for misleading the public about the risk of OxyContin and paid more than \$600 million in fines. Less than six months later, the company applied to the FDA for approval of a new reformulated version of OxyContin that included a chemical to make it more difficult to crush and misuse \parencite{oxymedreview2009}. Although not completely effective in reducing misuse, it was approved by the FDA and after August 2010 accounted for all OxyContin sales in the United States. Until 2016, with Mallinrockdt's Xtampza ER, Purdue was the only prescription opioid manufacturer to make abuse-deterrent oxycodone pills. The majority of all oxycodone sold over this time was generic oxycodone that remained abusable.\footnote{Many other companies attempted to make abuse-deterrent opioid pills at the same time, as shown in \citet{webster2009update}, but Purdue was the first to market. \citet{adler2018overview} and \citet{pergolizzi2018abuse} list other opioids with an ADF.}

Most research shows that OxyContin misuse fell following the reformulation. As described in \citet{Cicero2015}, although some users were able to circumvent the abuse-deterrent  formulation (ADF) to inject or ingest, the reformulation did reduce misuse. \citet{evans2019reformulation} finds that the reformulation coincided almost exactly with a structural break in aggregate oxycodone sales, which had previously been increasing. Shortly after the OxyContin reformulation was implemented, researchers began to notice illicit drug use moving towards other drugs such as heroin or generic oxycodone (\cite{cicero2012effect}, \cite{coplan2013changes}, \cite{Alpert2017}, \cite{evans2019reformulation}, \cite{havens2014impact}, \cite{cassidy2014changes}). Our paper extends the analysis of the impact of reformulation on opioid use by separately identifying the shifts in OxyContin and generic oxycodone misuse.

We build upon a rich literature that studies opioid misuse through surveys or analysis of the aggregated ARCOS reports. Surveys mostly polled either informants or users themselves (for details see \cite{inciardi2009black}). The best surveys have been of users in smaller samples at individual treatment facilities, like in \citet{hays2004profile} and \citet{sproule2009changing}. However, selection bias is a problem for surveying treatment facilities, as that is a specific subset of patients whose habits may be different from the overall drug-using population (particularly because they are seeking treatment). Some researchers have also used the quarterly ARCOS reports to study national trends in consumption, like in \citet{alpert2018supply}, \citet{mallatt2018effect}, and \citet{sairam2014assessment}. The quaterly ARCOS reports have no information on the market share of each brand of prescription opioid, thereby restricting any analysis to the aggregate level only. Our work is closely connected to the second set of papers, but we are able to leverage ARCOS's transaction level data to distinguish sales of OxyContin from generic oxycodone. 

This newly released ARCOS data allows us to make two methodological improvements. First, the literature treats the OxyContin reformulation as an exogenous shock at the state level. This assumption is problematic because each state’s dependency on OxyContin as well as exposure to the reformulation is the result of the state’s regulatory environment (\cite{alpert2019origin}). These regulatory factors could have an impact on how people react to the reformulation, and thus create a hidden link between OxyContin exposure and the reformulation outcomes. Using the new ARCOS data, we can disaggregate to Metropolitan Statistical Areas (MSAs), which allows our model to identify drug substitutions using within-state variations in opioid sales and mortality while controlling for across-state variations in policies and drug enforcement.

The second benefit of the new ARCOS data set is that it allows us to disaggregate different kinds of prescription opioid sales on a national scale.  Previous national studies were unable to distinguish between these drugs due to limitations in existing data. The NSDUH survey, the primary data source for drug misuse at the national level, only documented past year use of OxyContin. Death certificates do not distinguish between OxyContin and generic oxycodone. The aggregate ARCOS sales group all oxycodone sales into one bin. Because of OxyContin's unique role in fomenting the opioid epidemic, it has received most of the attention of researchers. The literature assumes that the study of OxyContin was equivalent to the study of all oxycodone.  As a result, although non-OxyContin oxycodone misuse is significant in size, it has been understudied. One notable exception is \textcite{paulozzi2006opioid}, which notes that non-OxyContin oxycodone was a better predictor of state opioid deaths than OxyContin.

The previous literature also attempts to link the misuse of prescription opioids to the rise in heroin misuse. \textcite{siegal2003probable} are the first to suggest the pathway from prescription opioids to heroin, and they further note a reverse in trend where heroin users switched to prescription opioids when heroin was unavailable. \textcite{compton2016relationship} describes how by the 21st century people who initiated heroin use were very likely to have started by using prescription opioids non-medically. The most recent works on OxyContin reformulation suggest that the reformulation played an important part in reigniting the heroin epidemic since 2010. \textcite{Cicero2015} and \textcite{mars2014every}, who rely on smaller surveys, find the predominant drug people switched to after reformulation was heroin. \textcite{evans2019reformulation} identifies a structural break in heroin deaths in August 2010 that was accompanied by higher growth in heroin deaths in areas with greater pre-reformulation access to heroin and opioids. Similarly, \textcite{alpert2018supply} shows that the rise in heroin deaths was larger in places with higher OxyContin misuse pre-reformulation. However, the evidence linking the reformulation to the rise in heroin death is not conclusive: other researchers suggest the sharp rise in heroin use may have predated the OxyContin reformulation by a few years (\cite{dasgupta2014observed}, \cite{cassidy2014changes}). With the new ARCOS data, we are able to examine the claim that the OxyContin reformulation caused the subsequent heroin epidemic in more detail. In particular, we separate the impact of the reformulation on heroin use from the gradual shifts in oxycodone misuse that are independent of the reformulation.

\section{Data and Descriptive Statistics}

To estimate the impact of the OxyContin reformulation on opioid use and mortality, we combine several data sources including sales of OxyContin and non-OxyContin alternatives from ARCOS, opioid and heroin mortality from the NVSS, and self-reported OxyContin and Percocet misuse from the NSDUH. Our main regression leverages variations in pre-reform exposure to OxyContin and generic oxycodone to identify the impact of the reformulation on opioid sales and mortality. We define a new measure of exposure by interacting the state-level self-reported opioid misuse and MSA-level opioid sales. In this section, we describe the three sources of data, the market definition, the construction of the OxyContin and generic oxycodone exposure measure, and present summary statistics of our data.

\subsection{Data}

\subsubsection{ARCOS and the Sales of Prescription Opioid}

As part of the Controlled Substances act, distributors and manufacturers of controlled substances are required to report all transactions to the DEA. This Automation of Reports and Consolidated Orders System (ARCOS) database contains the record of every pain pill sold in the United States. The complete database from 2006 to 2014 was recently released by a federal judge as a result of an ongoing trial in Ohio against opioid manufacturers.\footnote{\href{https://www.washingtonpost.com/graphics/2019/investigations/dea-pain-pill-database/}{Link} to the ARCOS Data published by the Washington Post.} 

The ARCOS database has been used previously to study opioids, but only using the publicly available quarterly aggregate weight of drugs sold (\cite{sairam2014assessment}) or via special request to the DEA (\cite{modarai2013relationship}). The newly released full database reports the manufacturer and the distributor for every pharmacy order. These data allow us to track different brands of prescription opioids separately, and calculate what fraction of oxycodone sold is OxyContin at any level of geographic aggregation. We can thus construct what we believe is the first public time-series of OxyContin and generic oxycodone sales from 2006-2014.

	\begin{figure} [H]
		\centering
	
		\begin{minipage}{0.9\textwidth}\
        \includegraphics[width=\linewidth]{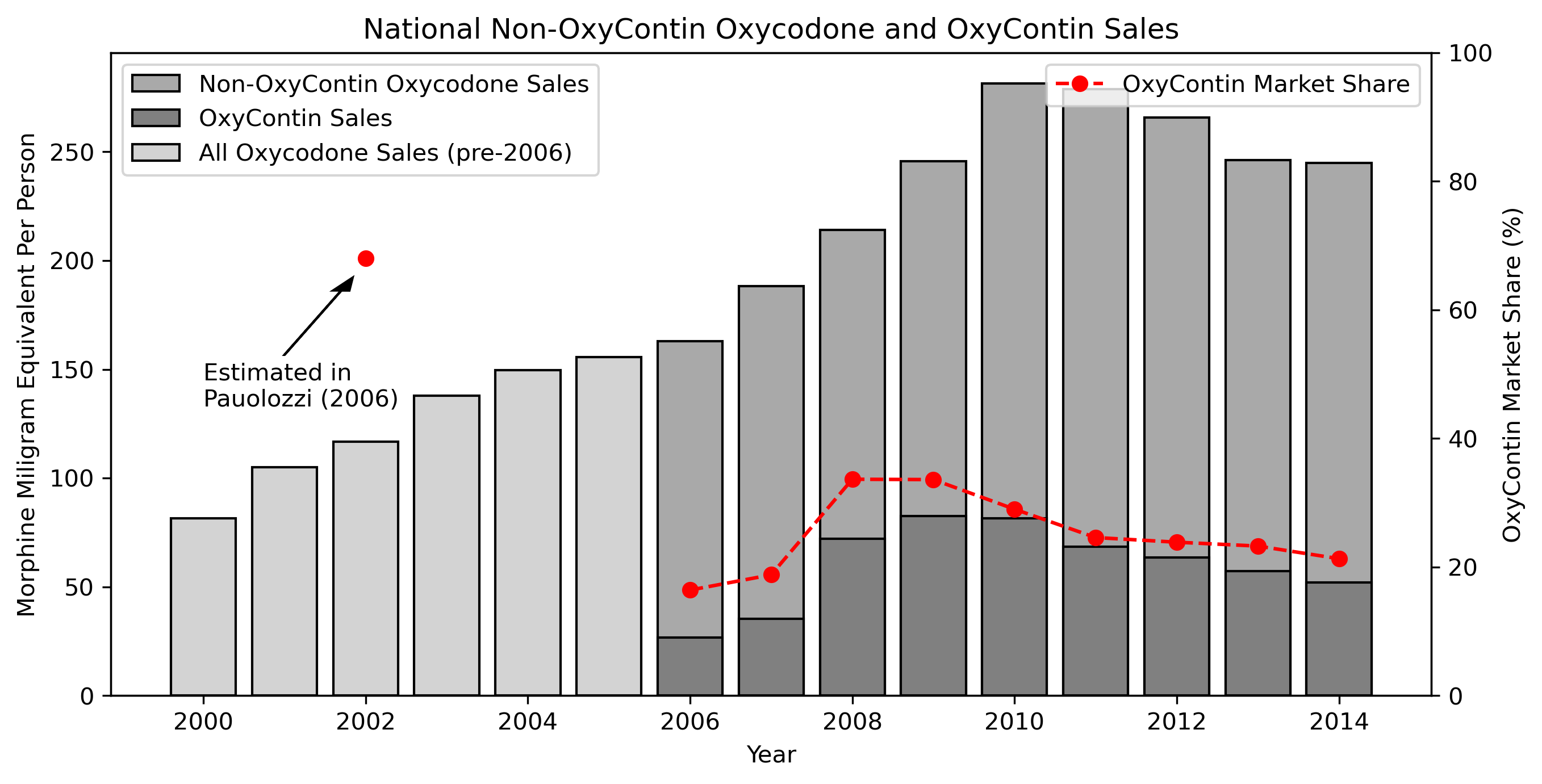}
        {\footnotesize \textit{Note:} We supplemented the 2006 to 2014 data with publicly available aggregate data from 2000 to 2005. The publicly available aggregate data does not break down the oxycodone sales by manufacturer. \par}
        \end{minipage}
		
		\caption{Growth of oxycodone and OxyContin sales}
		\label{fig:overall_growth}
	\end{figure}

As we can see from Figure \ref{fig:overall_growth}, total oxycodone sales increased substantially from 2000 to 2010, with per-person sales nearly quadrupling in the ten years period. From 2010 to 2015, sales of oxycodone declined as a result of aggressive measures taken by the states and the federal government to counter opioid addiction (\cite{kennedy2016opioid}). 

The newly available ARCOS data suggests that the commonly held belief about OxyContin’s dominance in the prescription opioid market at the time of reformulation is incorrect. The last time OxyContin’s market was estimated was in 2002 by \textcite{paulozzi2006opioid}, who acquired from the DEA a year's worth of ARCOS data aggregated at the state level. In that year, OxyContin was 68\% of all oxycodone sales by active ingredient weight and scholars have assumed that Purdue's market share stayed high until the OxyContin reformulation. However, as Figure \ref{fig:overall_growth} shows, by 2006 when our data starts, OxyContin sales only accounted for 18\% of all oxycodone sold by weight and never got above 35\% during this period. The share is even smaller if we count the number of pills sold, since the average OxyContin active ingredient weight is 5 to 10 times higher than that of oxycodone from other brands. The share of OxyContin decreased dramatically from 2002 to 2006 because Purdue lost the patent rights in 2004. As a result, non-OxyContin oxycodone sales grew much faster in the early 2000s than OxyContin sales. Figure \ref{fig:market_share} in Appendix presents the market share for all oxycodone manufacturers by dosage strength, and Purdue Pharma is only dominant at higher dosages ($\geq$ 40mg). The overestimation of OxyContin's importance in the pre-reform period explains why the previous literature overlooked the role generic oxycodone played in the opioid epidemic. 

The ARCOS sales data are the primary variables in our main regressions. We aggregate sales by MSA, year, and brand. To focus on the impact of the reformulation on OxyContin and non-OxyContin alternatives, we group all alternative oxycodone products into one measure, and we will refer to it as generic oxycodone for the rest of the analysis.\footnote{We acknowledge some non-OxyContin alternatives are branded and non-generic (i.e. Percocet and Percodan or later Roxicodone), but the majority of them are generic products. Generic oxycodone in this paper should be interpreted as all non-OxyContin oxycodone products.}

\subsubsection{NVSS Mortality Data}

The second outcome of interest in our main regression is opioid mortality. We use the restricted-use multiple-cause mortality data from the National Vital Statistics System (NVSS) to track opioid and heroin overdose. The dataset covers all deaths in the United States from 2006-2014. We follow the literature's two step procedure to identify opioid-related deaths. First, we code deaths with ICD-10 external cause of injury codes: X40–X44 (accidental poisoning), X60–64 (intentional self-poisoning), X85 (assault by drugs), and Y10–Y14 (poisoning) as overdose deaths. Second, we use the drug identification codes, which provide information about the substances found in the body at death, to restrict non-synthetic opioid fatalities to those with ICD-10 code T40.2, and heroin deaths to those with code T40.1. Figure 2 shows the trend over our period of study for the two series.

	\begin{figure} [H]
		\centering
	
		\begin{minipage}{0.9\textwidth}\
        \includegraphics[width=\linewidth]{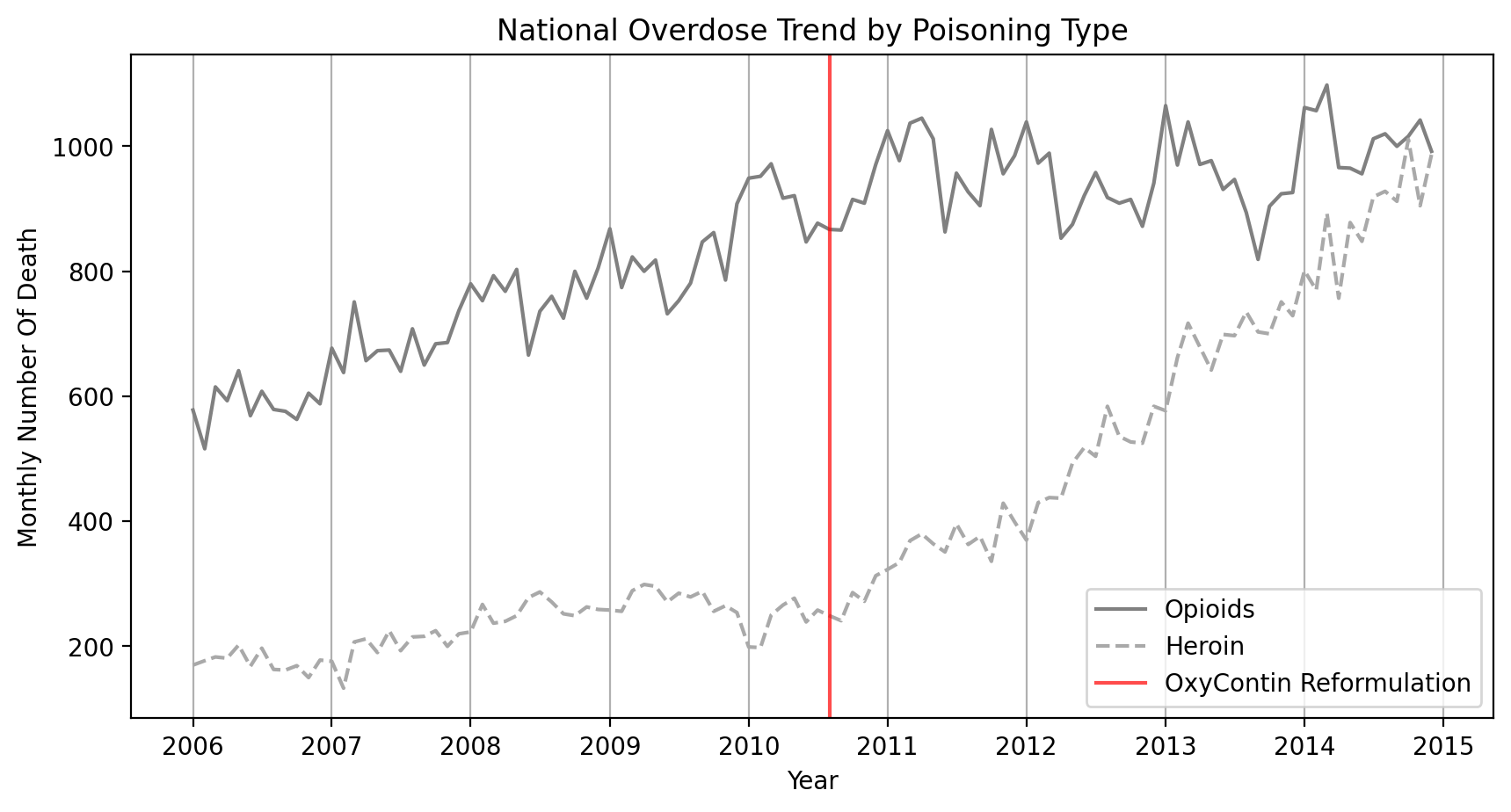}
        \end{minipage}
		
		\caption{Mortality trend}
		\label{fig:overdose}
	\end{figure}
	
The number of opioid fatalities grew in our sample period, from on average 600 deaths per month to 1000 per month. The number of heroin deaths was stable from 2006 to 2009 at about 200 deaths per month, and then it rose sharply from 2011 to 2015. As we've stated in the literature review section, the cause of the increase in heroin mortality is unclear. While some papers blame the OxyContin reformulation, there is evidence indicating the availability of heroin increased substantially after 2010 \parencite{o2017trends}.  

Since the number of drug overdose deaths with no drug specified accounts for between one-fifth and one-quarter of the overdose cases (\cite{ruhm2017geographic}), our measures of opioid and heroin deaths likely underestimate the true number of deaths.\footnote{Specifically, we omit ICD-10 code T50.9 (unspecified poisioning) from our analysis, and some fraction of these deaths are due to opioids or heroin but were not diagnosed or recorded as such.} However, the underestimation would not pose a problem for our regressions. There are variations in how coroners attribute the cause of death across states, but such variation would be captured by the state fixed effects. In addition, we do not anticipate systematic changes to each state’s practices due to the reformulation.

\subsubsection{NSDUH and Measuring Misuse}

We use state-level data from the National Survey on Drug Use and Health (NSDUH) to measure nonmedical use of opioids. The NSDUH publishes an annual measure of OxyContin misuse, asking the respondents whether they have ever used OxyContin ``only for the experience or feeling they caused" (\href{https://www.datafiles.samhsa.gov/sites/default/files/field-uploads-protected/studies/NSDUH-2006/NSDUH-2006-datasets/NSDUH-2006-DS0001/NSDUH-2006-DS0001-info/NSDUH-2006-DS0001-info-codebook.pdf}{NSDUH Codebook}). As first described in \citet{alpert2018supply}, the advantage of the NSDUH misuse measure is that it seperates out misuse from medical use. However, only OxyContin is reported in the NSDUH and there is no equivalent measure for generic oxycodone. 

Fortunately, the NSDUH reports PERCTYL2, which asks whether individuals ever misused Percocet, Percodan, or Tylox.\footnote{\href{https://www.accessdata.fda.gov/drugsatfda_docs/label/2006/040330s015,040341s013,040434s003lbl.pdf}{Percocet Drug Information.} Tylox was discontinued in 2012 following the \href{https://www.fda.gov/drugs/drug-safety-and-availability/fda-drug-safety-communication-prescription-acetaminophen-products-be-limited-325-mg-dosage-unit}{FDA regulations} limiting acetaminophen.} These drugs are oxycodone hydrochloride with acetaminophen and have a maximum dosage of 10mg of oxycodone per pill. The three drugs were popular among users in the pre-OxyContin era \parencite{meier2003pain}. In the present day, the PERCTYL2 variable captures misuse of not only the three branded drugs but also other generic oxycodone products that are popular on the street. 

The most direct evidence supporting this claim is the fact that generic oxycodone pills have often been referred to as `Percs' colloquially in the last decade. Many news report indicated that generic oxycodone has the street name `Perc 30' but is in fact not Percocet. The Patriot Ledger reported in a 2011 article\footnote{\href{https://www.patriotledger.com/article/20111219/NEWS/312199817}{Patriot Ledger Link.} Other references to generic non-OxyContin oxycodone as Perc 30s:  \href{https://www.phoenixhouse.org/news-and-views/true-stories/true-story-alex-2/}{Pheonix House}, \href{https://www.wsp.wa.gov/breathtest/docs/dre/manuals/inservice/2011/pharmageddon04.pdf}{Washington State Patrol}, \href{http://archive.boston.com/news/local/massachusetts/articles/2011/07/12/witness_police_work_led_to_arrest_of_musician_in_holdup/}{The Boston Globe}, \href{https://www.salemnews.com/news/local_news/dealers-life-goes-from-the-prep-to-state-prison/article_fccdc120-f578-5d92-9c3d-f6e797090edd.html}{The Salem News}, \href{https://www.mass.gov/doc/fitchburg-william-l-conlin-jr-dba-conlinscorner-violation-narcotics-09-10-15/download}{Massachusetts Court Filing}, \href{https://www.capecodtimes.com/article/20100912/NEWS/100909841}{Cape Cod Times}, \href{https://www.poconorecord.com/article/20110506/NEWS/105060380}{Pocono Record}, \href{https://bangordailynews.com/2014/05/09/news/bangor/key-witness-in-triple-murder-trial-arrested-for-violating-bail/}{Bangor Daily News}, \href{https://patch.com/massachusetts/tewksbury/dorringtons-we-did-it-to-protect-our-little-brother}{Patch}, \href{https://www.cnn.com/2011/OPINION/06/23/zeller.oxycodone.heroin/index.html}{CNN Op-Ed} } that `Perc 30s' were the newest drug of choice in South Shore of Massachusetts, saying:
\begin{quote}
\setlength{\leftmargin}{0.1in}
\setlength{\rightmargin}{0.1in}
\textit{`Perc 30s are not Percocet \textbf{\textemdash} the brand name for oxycodone mixed with acetaminophen, the main ingredient in Tylenol \textbf{\textemdash} but a generic variety of quick-release oxycodone made by a variety of manufacturers. They are sometimes referred to as ``roxys'' after Roxane Laboratories, the first company to make the drug, or ``blueberries,'' because of their color.'}
\end{quote}
Since many generic oxycodone users wouldn't know the name of the drug they use other than by its street name, but could distinguish between immediate release oxycodone and extended release OxyContin, it is likely that they answer affirmatively to misusing Percocet when they are, in fact, using generic oxycodone.\footnote{In the ARCOS dataset these pills are simply listed as `Oxycodone Hydrochloride 30mg'}

There are also several empirical observations that support this claim. The first is that we continue to see increases in the lifetime misuse of Percocet, Percodan, and Tylox even after they were replaced by OxyContin as the preferred prescription opioid to misuse. The misuse rate of Percocet, Percodan, and Tylox increased 30\% from 4.1\% to 5.6\% from 2002 to 2009 (see Figure \ref{fig:national_misuse} in Appendix), which would not have been possible if these drugs, or what people believed were `Percs', were not actively misused by new users post-introduction of OxyContin.

The second observation is that, based on the average sales data from 2006 to 2014, a disproportionate number of people has reported misusing Percocet, Percodan, or Tylox as compared to the actual sales of the three drugs. The sales of Endo Pharma, the manufacturer of Percocet and Percodan\footnote{Tylox not included since it was discontinued.}, are orders of magnitude less than the sales of Purdue while more than twice as many people reported misusing the three drugs as compared to OxyContin (see Figure \ref{fig:purdue_endo} in Appendix). A back-of-the-envelope calculation shows that if PERCTYL2 misuse captures only the misuse of Percocet and Percodan, then the proportion of pills misused out of all pills sold is 29 times higher for Percocet and Percodan than than the same proportion for OxyContin\footnote{In terms of number of pills circulated, OxyContin is 12.1 times Percocet and Percodan from 2006 to 2014. In terms of misuse, OxyContin is 41\% of Percocet and Percodan in the same period.}, an very unlikely situation given the popularity of OxyContin on the street.

This deduction is further supported by misuse data reported in the NSDUH. We know that generic oxycodone is commonly misused.\footnote{Law enforcement and journalists have previously identified the 30mg oxycodone pill as the most commonly trafficked opioid, see  \href{https://www.dea.gov/press-releases/2011/02/24/dea-led-operation-pill-nation-targets-rogue-pain-clinics-south-florida}{DEA Link},  \href{https://www.ice.gov/news/releases/10-facing-federal-drug-trafficking-charges-related-distribution-opioids-through-bogus}{ICE Link}, and \href{https://heroin.palmbeachpost.com/how-florida-spread-oxycodone-across-america/}{Palm Beach Post Link}.} If oxycodone has any other drug names, the popularity of that drug name in the NSDUH surveys should increase to reflect the increase in misuse in recent years. In addition to inquiring about popular brands, the NSDUH survey asks respondents to list any other prescription oxycodone that they have misused before. Dozens of pain relievers are reported, but in 2010 ``oxycodone or unspecified oxycodone products'' was only named by 0.10\%\footnote{\href{https://www.datafiles.samhsa.gov/sites/default/files/field-uploads-protected/studies/NSDUH-2010/NSDUH-2010-datasets/NSDUH-2010-DS0001/NSDUH-2010-DS0001-info/NSDUH-2010-DS0001-info-codebook.pdf}{NSDUH Codebook} variables ANALEWA through ANALEWE list the other pain relievers reported. Even if we assumed all 2.49\% of respondents saying they took a prescription pain reliever not listed had taken generic oxycodone, it is still less than half of the reported Percocet misuse.} of the respondents. No other brand of oxycodone pills are reported as commonly misused. We know from the reports in press and documents in court that generic oxycodone is a popular opioid on the street, and we know that Percocet is the only other commonly misused opioid documented in the NSDUH survey. Thus, the only way to reconcile the discrepancy between these two sources is that people mistakenly perceive generic oxycodone as Percocet or respond to the NSDUH as if they do. Thus, we use lifetime OxyContin and lifetime Percocet misuse for the construction of OxyContin and generic oxycodone exposure measures in Section 3.3.

\subsection{Market Definition and Endogeneity Problems}

Previous studies of the OxyContin reformulation depend on state-level variation to causally identify the impact of the reformulation. Treating OxyContin reformulation as an exogenous shock at the state level is potentially problematic. Although the timing of the reformulation is exogenous, each state’s exposure to it is a result of a combination of the state’s regulatory environment and Purdue’s initial marketing strategy (\cite{alpert2019origin}). These factors have substantial impact on how people in a state respond to the reformulation, creating a hidden link between exposure to the reformulation, the identifying variation, and subsequent drug use, the outcome variable.

One can limit the impact of endogenous regulation by disaggregation, but only if there is substantial intra-state variation in exposure to the reformulation. Both the ARCOS database and the NVSS mortality data have great geographic detail. Conducting our analysis on metropolitan statistical areas (MSAs), we find large variation in both OxyContin use and opioid mortality across MSAs in the same state. At the aggregate level in 2009, the average OxyContin market share in a state is 35.6\%. 65 of the 379 MSAs (17.1\%) in our sample have an OxyContin market share that is 10\% greater or smaller than their state average. The average opioid mortality is 0.343 deaths per 100,000 population in 2008. The variation in death is even more significant. More than 310 (83\%) MSAs have a mortality rate 20\% higher or lower than their state average, and more than 192 (51\%) have a mortality rate 50\% higher or lower than their state average . We present the full distribution of deviations of OxyContin market share and opioid mortality from state average in Figure \ref{fig:variation_oxy} and Figure \ref{fig:variation_mortality} in the Appendix.

Disaggregating to the MSA-level allows us to control for the state’s regulatory environment and hence eliminate the most problematic source of endogeneity. We use intra-state variation in exposure to the reformulation for identification. Intra-state heterogeneity in opioid use is associated with past economic conditions (\cite{carpenter2017economic}), location of hospitals and treatment centers (\cite{swensen2015substance}), preferences of local physicians (\cite{Schnell2017}), and local policy, some of which could still be correlated with the locality’s response to the reformulation. Analysis at the MSA level clearly allows us to make a much stronger claim than analysis at the state level. 

In addition, as we will show in the next sections, the disaggregation increases the statistical power of our regressions beyond the impact of the tripled sample size. Our results indicate that defining the market at the MSA level better captures the interaction between drug use and mortality than the state level. The important variations in drug use, for example between Los Angeles-Long Beach-Santa Ana at 4.4\% of nonmedical use of pain relievers and San Francisco-Oakland-Fremont at 5.6\%, disappears when they’re aggregated to the state level \parencite{nsduhMSA2012}.

\subsection{OxyContin and Non-OxyContin Oxycodone Exposure}\label{sec:misuse}


Since the OxyContin reformulation was a national event independent of local conditions, we can estimate its impact by comparing the outcomes in areas of high prior exposure to opioids with outcomes in areas of low exposure. Ideally, we want to quantify exposure using the volume of OxyContin misused in each region pre-reform while controlling for the volume of generic oxycodone misused. In practice, we do not observe these quantities. The best proxy in the literature is the self-reported misuse rate from the NSDUH.

Based on the NSDUH misuse, we create a new measure of OxyContin and non-OxyContin oxycodone exposure by combining the NSDUH state-level misuse rate with ARCOS MSA-level sales. Specifically, for each drug, we calculate:
\begin{equation}
\text{Exposure}_{m}^{\text{pre-reform}} = \text{Lifetime Misuse}_{s}^{2004-2009} \times \text{Sales}_{m}^{2009}
\end{equation}
Our measure is the interaction term of sales of OxyContin/generic oxycodone in an MSA and the lifetime misuse rate of that drug in the corresponding state. This new measure has two advantages over the conventional misuse rate from NSDUH: it captures intra-state variation in misuse and it more accurately reflects the current misuse of both OxyContin and generic oxycodone.

The NSDUH surveys approximately 70,000 respondents every year and uses sophisticated reweighting techniques to get accurate state level estimates. Once we get to the MSA level, the number of people surveyed as well as the number of positive responses to questions on opioid misuse are extremely small. As a result, most of the outcomes at the MSA level are censored by the NSDUH to protect individual privacy. Using only the survey data means that we would use same state misuse value for all MSAs and therefore forgo any intra-state variation in drug use. In comparison, our proposed measure relies on deviations from normal sales patterns to generate variations in exposure rates for the MSAs. Our definition assumes that the percentage of people reported misusing a particular drug in a state is equivalent to the proportion of sales that are being misused. In a state where all the MSAs have identical sales, all the MSAs will have identical exposure rates by definition. However, if one MSA has higher sales of OxyContin compared with the rest of the state, our OxyContin exposure measure in that MSA will be higher than the rest of the state. This construction of exposure mirrors our intuitive understanding that the misuse of a drug in a locality is a function of the overall misuse and the availability of that particular drug in the area.

The NSDUH survey\footnote{In all surveys prior to 2014.} reports past-year misuse of OxyContin but only lifetime misuse of generic oxycodone. Previous studies did not focus on generic oxycodone misuse, so these studies rely on past-year OxyContin misuse rate. In our case, to disentangle substitution among prescription opioids, we have to make the comparison between OxyContin and generic oxycodone equal. Resorting to lifetime misuse rates for both series sacrifices the timely nature of the NSDUH misuse rates. By combining the lifetime misuse rates with sales in the year before reformulation, we capture recent changes in use of both drugs. To make our results comparable with previous studies, in the Appendix section, we repeat our entire analysis with OxyContin last-year misuse and generic oxycodone lifetime misuse. Most of our conclusions stand despite giving OxyContin a more favorable treatment. 

To construct our measure, we follow the precedent set in the literature by using a six-years average state level lifetime misuse rate pre-reform (2004 - 2009) and sales in 2009. The goal of the time average is to reduce the variance of the state-level misuse rates. We check the validity of our measure by regressing opioid death on it and compare the results with the same regressions on either only ARCOS sales or only NSDUH misuse. Results are summarized in Table \ref{tab:justification} in Appendix. The fit of the generic oxycodone regression is much improved with the interacted variable ($R^2 = 0.187$) relative to using only one with NSDUH misuse ($R^2 =0.062$) or sales ($R^2 = 0.176$). The improvement is even larger for the OxyContin regression ($R^2 = 0.128$) relative to using only one with NSDUH ($R^2 = 0.084$) or with sales ($R^2 = 0.086$).

\subsection{Descriptive Statistics}

\begin{table}
\footnotesize
\renewcommand\arraystretch{1.1}
\begin{threeparttable}
\input{table2_summary_stat}
\end{threeparttable}
\end{table}

Table \ref{tab:table2} reports summary statistics for five groups of MSAs: All MSAs, MSAs with high OxyContin exposure, MSAs with low OxyContin exposure, MSAs with high generic oxycodone exposure, and MSAs with low generic oxycodone exposure. MSAs with high OxyContin exposure and MSAs with high generic oxycodone exposure have similar demographic summary statistics. These two groups of MSAs also are not different statistically in their heroin mortality. Disentangling the impact of various opioids on the rise in heroin mortality is impossible with nationally aggregated or state level data due to the high correlation in misuse. The high correlation also implies that regressing heroin death on OxyContin without controlling for generic oxycodone use will likely lead to an overestimation of OxyContin's impact.

MSAs with high misuse differ from MSAs with low misuse. High misuse states have higher sales of both types of prescription opioids (twice as much for both types of opioids), higher mortality rate (twice as much for both opioid and heroin overdose), smaller population, higher average age, higher median income, higher percentage of white population, and lower percentage of black population. The differences in racial composition repeat well established findings in the literature: prescription opioid misuse was originally concentrated among white users, and by 2010 new heroin users were almost entirely white \parencite{cicero2014changing}. These differences in demographic variables motivate the inclusion of control variables in our main regressions.

\section{Empirical Strategies}

Our goal is to investigate two questions. First, what was the reformulation's immediate impact on OxyContin and generic oxycodone use? Second, what was the reformulation’s long-run effect on opioid mortality, heroin mortality, and on the progression of opioid addiction? 

We follow the event study framework from \textcite{alpert2018supply} to estimate the causal impact of the OxyContin reformulation on OxyContin and generic oxycodone sales and opioid and heroin mortality. We exploit variation in MSAs’ exposure to the reformulation due to the differences in their pre-reform OxyContin use while controlling for pre-reform generic oxycodone use. Our approach is similar to \citet{finkelstein2007aggregate}, where the OxyContin reformulation has more ``bite'', or more of an effect, in areas where OxyContin misuse was higher than in places where generic oxycodone was the preferred drug. The approach allows us to measure whether MSAs with higher exposure to OxyContin experienced larger declines in OxyContin sales, larger increases in alternative oxycodone, or larger increases in opioid and heroin mortality. The empirical framework is:

\begin{equation}
\label{reg1}
\begin{split}
Y_{mt} &= \alpha_{s} + \delta_{t} + \sum_{i=2006}^{2014} \mathbbm{1}\{i = t\} \beta_i^1 \times \text{OxyContin Exp}^{Pre}_{m} \\
+& \sum_{i=2006}^{2014} \mathbbm{1}\{i = t\} \beta_i^2 \times \text{Oxycodone Exp}^{Pre}_{m} + X'_{mt}\gamma +\epsilon_{mt}
\end{split}
\end{equation}

where $Y_{mt}$ are the outcome variables of interest in MSA $m$ at year $t$; $\text{OxyContin Exp}^{Pre}_{m}$ and $\text{Oxycodone Exp}^{Pre}_{m}$ are time-invariant measures of OxyContin and oxycodone exposure before the reformulation (see Section 3.5 for construction), and are interacted with a set of $\beta_t^1$ and $\beta_t^2$ for each year. We include state fixed effects to control for regulatory differences among states and year fixed effects to control for national changes in drug use. We also include a full set of MSA-level demographic variables. We weight the regression by population and exclude Florida.\footnote{The literature excludes Florida because it underwent massive increases in oxycodone sales over this period, some of which was trafficked to other states.} We show the full set of $\beta_t$ estimates graphically, normalizing by the 2009 coefficient. The $\beta_t$ identifies the differences in sales and death across MSAs due to their higher or lower pre-reform OxyContin or oxycodone exposure. Standard errors are clustered at the MSA level to account for serial correlation. In the Appendix section, we present beta estimations from variations of our base model, which include (1) using a MSA fixed effect instead of state fixed effect, (2) replacing OxyContin lifetime misuse rate with OxyContin last-year misuse rate, (3) regressing at the state level, and show that our conclusion are insensitive to most of these variations.

To complement our results, we also use a strict difference-in-difference framework to estimate effect of the reformulation conditioning on OxyContin and non-OxyContin oxycodone exposure levels. Our specification is:
\begin{equation}
\label{reg2}
\begin{split}
Y_{mt} &= \alpha_s + \gamma_t + \delta_1\mathbbm{1}\{t > 2010\} \\
&+ \delta_2\mathbbm{1}\{m \in HighOxyContin\} + \delta_3\mathbbm{1}\{m \in HighOxycodone\} \\
&+ \delta_4\mathbbm{1}\{t > 2010\}\times \mathbbm{1}\{m \in HighOxyContin\} \\
&+ \delta_5\mathbbm{1}\{t>2010\}\times \mathbbm{1}\{ m\in HighOxycodone\} + X'_{mt}\beta + \epsilon_{mt}
\end{split}
\end{equation}
where $HighOxyContin$ and $HighOxycodone$ are the set of MSAs with higher than median pre-reform exposure to OxyContin and oxycodone respectively. We restrict the regression to include only the three years prior (2008 to 2010) and the three years after (2011 to 2013) the reformulation. The advantage of this specification is that it does not assume that OxyContin or oxycodone exposure affects the outcome variable linearly. Instead of having a flexible $\delta$ for each year, we have only one $\delta$ for each of the pre- or post-reform period. In this specification, we simply test whether higher exposure MSAs reacted differently to the reformulation as compared to lower exposure MSAs (if $\delta_4$ and $\delta_5$ are significant). We include state fixed effects to control for state-level heterogeneity, year fixed effects for national trend, and a set of time-varying MSAs level covariates. Again, standard errors are clustered at the MSA level.

\section{Results}

We proceed in two steps. First, we provide direct evidence that the OxyContin reformulation caused OxyContin sales to decrease and generic oxycodone sales to increase, and that the changes in sales are proportional to the pre-reformulation level of OxyContin exposure. Second, we estimate the impact of the reformulation on opioid and heroin mortality. We find that high pre-reformulation levels of OxyContin exposure were not associated with high opioid deaths, but there was a strong positive effect from generic oxycodone exposure in both the pre- and post-reform period. We find that higher pre-reform OxyContin and pre-reform oxycodone exposure were both positively but not significantly associated with later heroin deaths, but the oxycodone coefficient is larger. If we run the heroin regression separately with only OxyContin exposure we recover the results of the literature, but running the heroin regression with only oxycodone exposure better fits the data.  


\subsection{Reformulation's Impact on Opioid Sales}

\begin{figure} [H]
		\centering
	
		\begin{minipage}{0.8\textwidth}\
        \includegraphics[width=\linewidth]{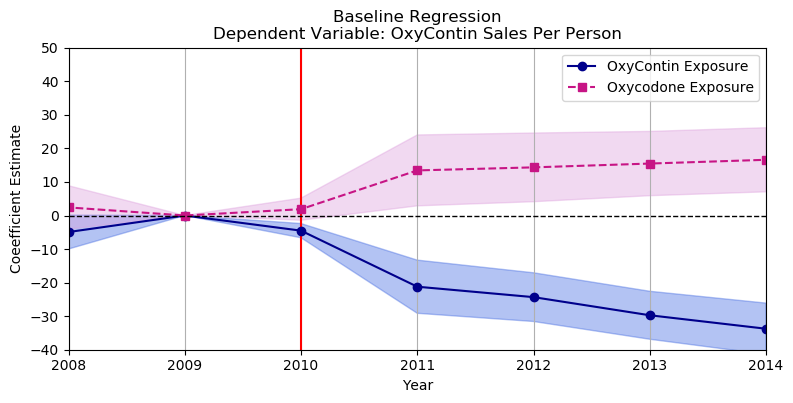}
        \end{minipage}
		
		\caption{Main regression on OxyContin sales. Shaded regions are the 95 percent confidence intervals with standard errors clustered at the MSA level.}
		\label{fig:main_oxycontin}
	\end{figure}

\begin{figure} [H]
		\centering
	
		\begin{minipage}{0.8\textwidth}\
        \includegraphics[width=\linewidth]{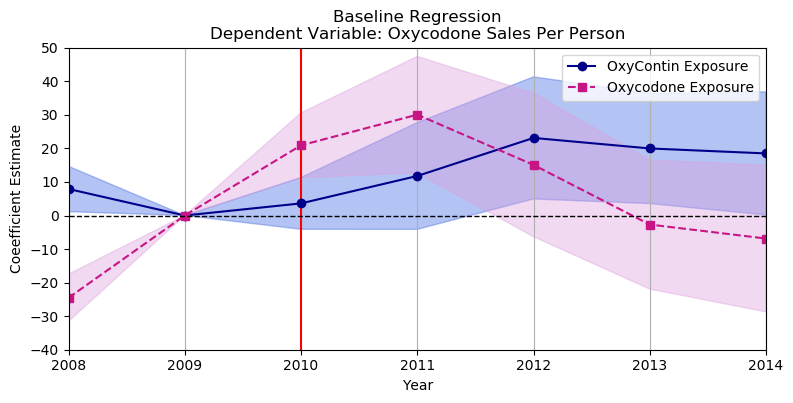}
        \end{minipage}
		
		\caption{Main regression on generic oxycodone sales. Shaded regions are the 95 percent confidence intervals with standard errors clustered at the MSA level.}
		\label{fig:main_oxycodone}
	\end{figure}

We begin by showing graphically that OxyContin sales decreased and generic oxycodone sales increased in high OxyContin misuse MSAs immediately after reformulation. Figure \ref{fig:main_oxycontin} and Figure \ref{fig:main_oxycodone} present the full set of coefficients from estimating the event study framework on OxyContin and generic oxycodone sales. Each data point in the figure is the coefficient of the interactive term of misuse and sale, which we call exposure, for OxyContin or generic oxycodone in a specific year, and it captures any additional change in sales in that year driven by high OxyContin or oxycodone exposure. In Figure \ref{fig:main_oxycontin}, we observe a larger decrease in OxyContin sales post-reform in MSAs with higher pre-reform OxyContin exposure. As Figure \ref{fig:main_oxycodone} shows, higher OxyContin exposure MSAs saw greater increases in generic oxycodone sales post-reform. The effects are well identified at 95\% confidence level. An one standard deviation increase in OxyContin exposure translates into an additional 21.2 MME decrease in per person OxyContin sales and 11.8 MME increase in per person oxycodone sales in 2011. These changes represents a 24\% decrease in OxyContin sales and a 8.8\% increase in oxycodone sales from the 2009 level. The effects are economically significant especially given that the reformulation should only affect the population abusing OxyContin, so this drop in sales is driven by a fraction of all users. The two observations combined support the hypothesis that reformulation caused substantial substitution from OxyContin to generic oxycodone.

Figure \ref{fig:main_oxycontin} also documents that high pre-reform oxycodone misuse MSAs saw large increases in OxyContin sales right after the reformulation. This phenomenon has been unreported previously, but would be consistent with \citet{Schnell2017}'s physician benevolence hypothesis where good physicians switch patients from oxycodone to reformulated OxyContin to lower the future risk of abuse. Although the switch toward OxyContin is smaller in magnitude than the switch from OxyContin, this increase is the first documented positive impact of the OxyContin reformulation in the literature. It seems both physicians and users saw the two types of drugs as substitutes. Unfortunately, there are not enough MSAs where the switch toward OxyContin is significant enough that it cancels out the switch away from OxyContin to examine the possible substitution channel in the other direction.

Because we include both OxyContin and generic oxycodone misuse in the same regression, we can separate out the increases in oxycodone sales due to its own popularity from the increases due to spillover effects from the OxyContin reformulation. Figure \ref{fig:main_oxycodone} shows increasing growth in oxycodone sales in MSAs with higher oxycodone misuse until 2011, and the growth rate declined after. The smoothness of the oxycodone curve indicates that the OxyContin reformulation had no impact on how oxycodone misuse affected oxycodone sales. This trend corresponds well with many states tightening control over opioid prescription policies in 2011 and 2012 in response to rising sales and and increased awareness of opioid misuse. 

Another way of estimating the impact of the reformulation is through difference-in-difference regressions. Column (1) of Table \ref{tab:did} in Appendix shows the regression on OxyContin sales. OxyContin sales in all MSAs decreased by 8.05 MME post-reform, a 9.4\% decrease with respect to the average per person sales of 85.6 MME in 2009. High OxyContin misuse MSAs had a higher level of OxyContin sales to start with, but experienced an additional 15.1 MME drop (an additional 17\% decrease) post-reform. Given that only 2.46\% of the population ever misused OxyContin\footnote{NSDUH, 2010.} and the reformulation only affected the people misusing it, a 17\% additional decrease in all OxyContin sales would translate into a very significant decrease in sales to the population that misuses it. The negative and significant \textit{Post $\times$ High OxyContin} coefficient confirms previous findings that high OxyContin exposure MSAs saw larger decreases in OxyContin sales post-reform. 

Column (2) of the same table reports the regression on generic oxycodone sales. Generic oxycodone sales per person increased 41.7 MME in the post period, a 31.2\% increase with respect to the average per person alternative oxycodone sales of 133.5 MME in 2009. High OxyContin misuse MSAs experienced an additional 10.3 MME increase, which translates to a 68\% conversion from OxyContin to generic oxycodone in those areas. Combining the findings from columns (1) and (2), we see direct substitution from OxyContin to generic oxycodone in local sales immediately after reformulation, and the substitution pattern is more pronounced in MSAs with high OxyContin exposure as expected.

To help our readers visualize the trend of OxyContin and alternative oxycodone sales, in Figure \ref{fig:drop_1} in the Appendix, we break all MSAs into three bins by the magnitude of the observed drop in OxyContin sales due to the reform. Then, we plot the per person OxyContin and generic oxycodone sales for the three group respectively. By definition, the high empirical drop group experienced the largest decreases in OxyContin sales from 2009 to 2011 (-29\%) and the low drop group experienced an increase in OxyContin sales (+15\%). Sales of generic oxycodone started at different levels, but shared the same growth rate until the reformulation in 2010. Since 2010, the higher the empirical drop in OxyContin, the faster the growth in generic oxycodone. The high group saw in a 72 MME increase (46\% from 2009) in generic oxycodone sales, while the low group only saw an 29 MME increase (29\% from 2009).  The high growth rate of generic oxycodone in high drop MSAs support the substitution story. The post-reform level of OxyContin sales converges to the same level for all three groups, suggesting that the remaining sales most likely represent non-replaceable demand for medical OxyContin use. 

\subsection{Reformulation and Opioid and Heroin Mortality}

\begin{figure} [H]
		\centering
	
		\begin{minipage}{0.8\textwidth}\
        \includegraphics[width=\linewidth]{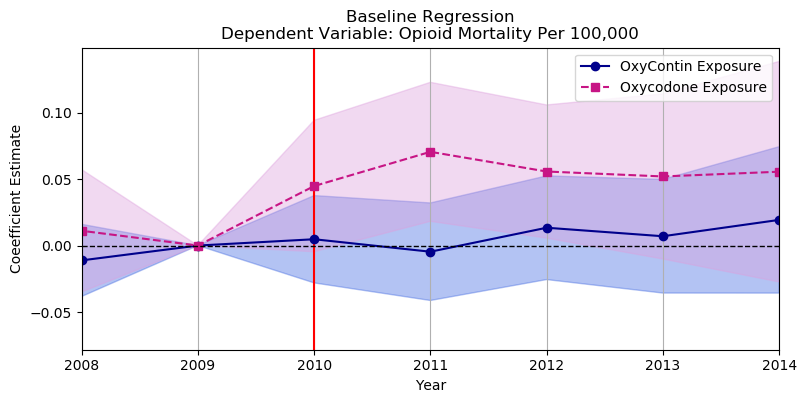}
        \end{minipage}
		
		\caption{Main regression on opioid mortality. Shaded regions are the 95 percent confidence intervals with standard errors clustered at the MSA level.}
		\label{fig:main_opioid}
	\end{figure}

Next, we estimate the impact of the reformulation on overdose mortality. In Figure \ref{fig:main_opioid}, we report the full set of coefficients from estimating the event study framework on opioid mortality. Each data point in the figure is the coefficient of the interactive term of misuse and sale for OxyContin or generic oxycodone in a specific year, and it captures any additional change in opioid mortality in that year driven by high OxyContin or oxycodone exposure. The OxyContin coefficients are never significant, suggesting higher pre-reform OxyContin misuse is not predictive of either higher or lower opioid death post-reform. The lack of any trend indicates that any benefit of the OxyContin reformulation on reducing OxyContin consumption is offset by the substitution to generic oxycodone. In aggregate, the reformulation had no impact on non-heroin opioid deaths. 

	\begin{figure} [H]
		\centering
	
		\begin{minipage}{0.8\textwidth}\
        \includegraphics[width=\linewidth]{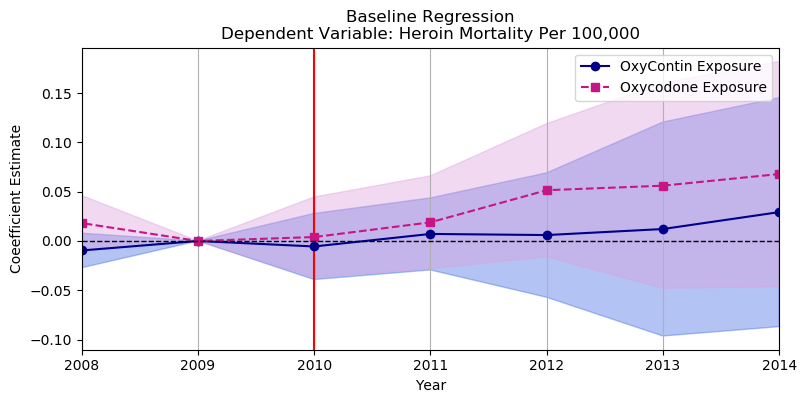}
        \end{minipage}
		
		\caption{Main regression on heroin mortality. Shaded regions are the 95 percent confidence intervals with standard errors clustered at the MSA level.}
		\label{fig:main_heroin}
	\end{figure}

In Figure \ref{fig:main_heroin}, we report the event study coefficients on heroin mortality. Again the OxyContin coefficients are tiny and insignificant, while the oxycodone coefficients grow over time but never reach statistical significance at conventional levels. The lack of statistical significance is due to the small number of heroin moralities in the whole sample and high correlations between OxyContin and oxycodone exposure. If we were to run the OxyContin and oxycodone regression separately (See Figure \ref{single_heroin1} and Figure \ref{single_heroin2} in Appendix), oxycodone exposure had a much larger and more significant impact on heroin mortality. The results provide tentative evidence that the higher the generic oxycodone exposure in an MSA, the higher the increases in heroin mortality. However, the results do not support the alternative hypothesis that the OxyContin reformulation was solely responsible for the increase in heroin mortality. 

The difference-in-difference results mirror our finding from the event study framework. Column (3) of Table \ref{tab:did} in Appendix suggests that opioid deaths are 0.08 higher in high oxycodone exposure MSAs, which is equivalent to 27\% of the average opioid overdose of 0.29 per 10,000 people in 2009. Opioid mortality is 0.05 lower (17\% of the 2009 average) in higher OxyContin exposure MSAs after controlling for oxycodone use. Higher OxyContin exposure does not lead to higher or lower opioid overdose post-reform, while higher 
generic oxycodone exposure is associated with 0.06 (20.6\% of 2009 average) more opioid death in the post period. 

Column (4) of the same table reports the difference-in-difference regression on heroin death. Heroin mortality has increased by 0.14 in the post period in all MSAs, which is equivalent to a 111\% increase from the average 2009 level of 0.126 heroin death per 10,000 population. High OxyContin exposure MSAs did not experience additional jumps in heroin mortality, while high oxycodone exposure MSAs did experience an additional 0.07 (56\% with respect to 2009 average) increase in death. Again, the evidence from the difference-in-difference regressions indicate that OxyContin was not responsible for the rise in heroin mortality.

In Figure \ref{fig:drop_2} in the Appendix, we show the average trend of the opioid and heroin mortality for groups with high, medium and low observed drop in Oxycontin sales. If the reformulation was responsible for the subsequent heroin epidemic, then the MSAs mostly likely to have additional jumps in heroin mortality would be the MSAs with the largest OxyContin drop. As shown in the figure, the three groups went through the same explosive growth in heroin mortality (around 38\% from 2009 to 2011, and similar rate afterward), indicating the rise in heroin was independent of the decrease in OxyContin sales. This evidence conclusively rejects the hypothesis that the OxyContin reformulation is solely responsible for the subsequent heroin epidemic.

\subsection{Discussion} \label{sec:discussion}

\textit{(A) The Reformulation's Impact on Opioid Mortality}

Until now, the literature has found mixed results for the effects of the OxyContin reformulation on opioid mortality. In contrast to previous work, we find no statistically significant impact of the reformulation on opioid mortality as a result of substantial substitutions from OxyContin to generic oxycodone post-reform. Increases in generic oxycodone sales
compensated for 55\% of the drop in OxyContin sales in high OxyContin misuse MSAs by our event study framework, and 68\% by our different-in-difference estimation. Opioid mortality continued to increase in the post-reform period, but not was driven by high OxyContin exposure. 


\textit{(B) The Reformulation's Impact on Heroin Mortality}

Our results stand in direct contrast to the findings of the literature. Instead of being the event that precipitated the heroin epidemic, the OxyContin reformulation shifted misuse to other opioids, of which heroin was only one. We cannot refute the hypothesis that some OxyContin users switched to heroin due to the reformulation. Our analysis refutes the hypothesis that the reformulation was the sole cause of the heroin epidemic. Instead of OxyContin misuse, we identified generic oxycodone misuse as a much more powerful driver of increases in heroin mortality post-2011. What prompted the increases in heroin use is still an unresolved question. Previous research has suggested an increase in the supply of heroin (\cite{o2017trends}) around this time, as well as crackdowns in Florida on pill-mills reducing the supply of oxycodone (\cite{kennedy2016opioid}).  

\textit{(C) Bridging the Differences between our Findings and the Literature}

One of the innovations we've made in this paper is to shed light on a hidden source of opioid misuse: the misuse of generic oxycodone. This segment of prescription opioids was overlooked by other scholars because of OxyContin's dominance in opioid misuse in the early years as well as, we argue, the lack of identifiable brand names for the generic products. Empirical studies based on market data or interviews of opioid users noted that many people misused generic oxycodone products (\cite{paulozzi2006opioid}, \cite{inciardi2009black}). Leaving out oxycodone misuse, an important driver of opioid and heroin mortality that is positively correlated with OxyContin misuse, would produce spurious regression results. 

To show that the difference in findings is not driven by our constructed misuse measure, or our choice of framework, we test whether we can reproduce findings in the literature by running all of our regressions using only OxyContin (see Section \ref{sec:only_oxy} in the Appendix). Our OxyContin misuse exposure individually predicts an increase in opioid and heroin mortality post-reform as the literature claims. This finding is the basis of previous studies supporting the claim that the OxyContin reformulation is the main cause of the subsequent heroin epidemic. However, if we run the same set of regressions using only generic oxycodone (see Section \ref{sec:only_codone}), we were able to produce the same findings. The only way to differentiate the impact of OxyContin from that of generic oxycodone is to include both in the same regressions. Variations in local OxyContin and oxycodone exposure allow us to identify the impact of both series, if any exist. As we've shown in our main regressions, the impact of OxyContin on heroin disappears after controlling for the effect of generic oxycodone. 

\textit{(D) Market Definition}

Another innovation we've made in this paper is a finer definition of the opioid market. It is important to consider what we gain from disaggregating to the MSA level. The specific OxyContin market share in a state is endogenous to a great many things, including advertising (\cite{van2009promotion}) and triplicate status (\cite{alpert2019origin}). Although the OxyContin reformulation was an exogenous shock, its interpretation is made very complicated because its impact depended on each state's regulatory history and prescribing environment. We do our regressions at the MSA level, where there are unobserved local conditions that affected sales of OxyContin and generic oxycodone, while controlling for state-level laws and restrictions. By comparing two different MSAs with the same regulatory environment but different exposures to the reformulation, we can get at the marginal effects of OxyContin and generic oxycodone exposure. Contrasting the state-level regression estimates (see Section \ref{sec:state}) with our main results, our main results are larger in magnitude and more statistically significant. The MSA level estimation of the effect of exposure on mortality is more stable.  



\textit{(E) Definition of OxyContin Misuse}

The literature relies on NSDUH's OxyContin past-year misuse. To make our findings comparable with previous studies and robust to the choice of misuse measure, we repeat our entire analysis with OxyContin last-year misuse and generic oxycodone lifetime misuse (see Section \ref{sec:OXYYR} for results.) As noted in Section \ref{sec:misuse}, using last-year OxyContin misuse gives an unfair advantage to OxyContin due to the timeliness of the measure. If our findings on oxycodone persist despite the unequal treatment of the two misuse measures, then it is a stronger indication of the essential role generic oxycodone played in the opioid and heroin epidemic. 

Comparing the two sets of results, we observe the same decline in OxyContin sales and increase in generic oxycodone sales, although smaller in magnitude. Both sets of coefficients on opioid mortality become positive but insignificant. Finally, comparing the heroin result, at the state level we do detect a positive effect on heroin mortality from OxyContin. In aggregate, our results lose some significance when we replace lifetime OxyContin misuse with last-year OxyContin misuse. The loss of significance, however, is in the direction predicted by the unfair advantage given to OxyContin. This exercise highlights the importance of treating the two misuse measures equally. When we use measures that more accurately capture recent OxyContin misuse than recent generic oxycodone misuse, we could mistakenly attribute effects of generic oxycodone to OxyContin.

\section{Conclusion} \label{sec:conclusion}

Researchers have attributed the prescription opioid opioid crisis and recent increase in heroin use to OxyContin. Previous studies have documented how Purdue Pharma's marketing downplayed the risks of OxyContin's abuse potential, which fomented the prescription opioid crisis; recent studies identified the OxyContin reformulation as the event that pushed users to switch to heroin, which precipitated recent increase in heroin use. This paper revisits the roles OxyContin and the Oxycontin reformulation played in the opioid crisis with fine-grained sales data that includes OxyContin's most immediate substitute, generic oxycodone. We have three main findings. 

First, we find direct evidence of substitution to from OxyContin to generic oxycodone post-reformulation. Our difference-in-difference estimation indicates a 68\% substitution from OxyContin to generic oxycodone due to the reform. Looking at the decline in OxyContin sales and rise in generic oxycodone sales from 2002-2006, we believe this substitution (for different reasons, namely Purdue's loss of its patent) also happened years before the reformulation. The size of this substitution, and indeed the size of the generic oxycodone market pre-reform, may come as a surprise to researchers. \cite{paulozzi2006opioid} estimate that in 2002 OxyContin's market share was 68\%. By the time of the reformulation in 2010, it had fallen by more than half. OxyContin played an essential part in igniting the prescription opioid crisis but, after losing its patent in 2004, other companies took up the torch and surpassed Purdue by selling generic oxycodone. 

Our second main finding is that the OxyContin reformulation had no overall effect on opioid mortality. In our estimation, the OxyContin coefficients are not significant in the entire sample period, suggesting that higher OxyContin exposure is not predictive of either higher or lower opioid death. The lack of any trend indicates that the benefits of the OxyContin reformulation, if exist, are offset by substitution to oxycodone. In addition, we do find that high oxycodone exposure is predictive of rise in opioid mortality from 2011, confirming the increasingly important role of generic oxycodone in the recent prescription opioid crisis. 

Third and most importantly, we show that the heroin overdose deaths after 2010 were predicted by generic oxycodone exposure, not OxyContin exposure. Our main event-study model shows positive and significant effects from oxycodone exposure on heroin deaths after 2012, but OxyContin exposure is not predictive of heroin deaths once we control for oxycodone. The difference-in-difference results are similar, showing that oxycodone exposure was predictive of heroin deaths before or after the reformulation, and OxyContin exposure after the reformulation is weakly positive but not statistically significant. We also do not observe an additional rise in heroin deaths immediately after reformulation in areas where OxyContin sales declined the most post-reformulation. In particular, without including generic oxycodone in the analysis, we recover the same results from the literature that OxyContin was responsible for the rise in heroin deaths. The evidence shows that omitting oxycodone, an important substitute to OxyContin, produces erroneous results. This paper demonstrates the pernicious effects of generic oxycodone, which had thus far escaped scrutiny until the Washington Post acquired data and reported on it.

\newpage
\printbibliography

\section{Appendix}
\subsection{Additional Information}

	\begin{figure} [H]
		\centering
	
		\begin{minipage}{0.9\textwidth}\
        \includegraphics[width=\linewidth]{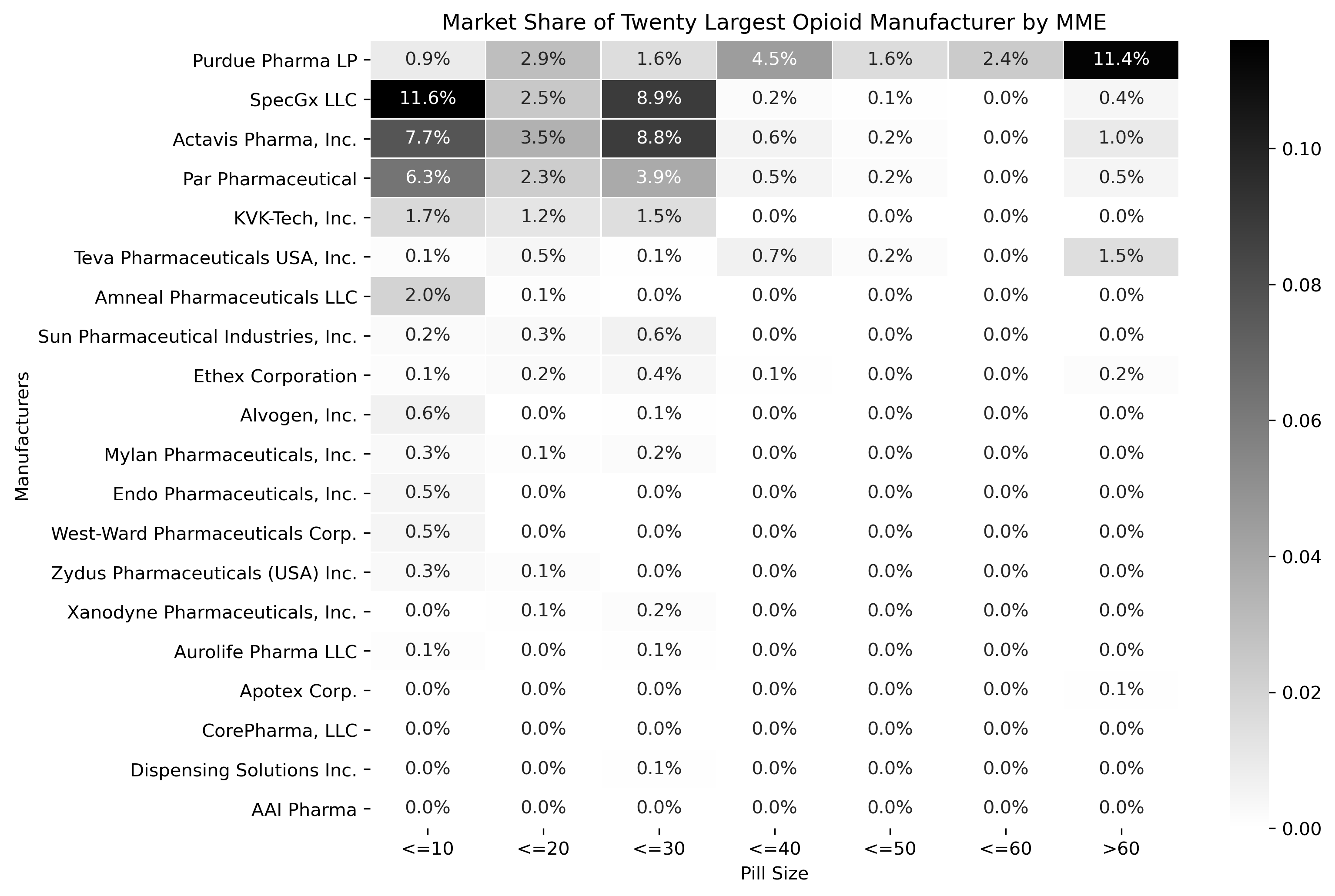}
        {\footnotesize \textit{Note:} We compute market share based on the average of 2006-2014 sales data. We kept only the top twenty manufacturers for better readability of the table. The rest of the 35 manufacturers combined contribute 0.18\% of total sales. During this sample period, Purdue Pharma was the dominant manufacturer of high dosage oxycodone pills ($\geq$ 40mg). In the lower dosage market, three manufacturers (SpecGx, Actavis Pharma and Phar Phamaceutical) had higher share of the market than Purdue Pharma. \par}
        \end{minipage}
		
		\caption{Market share of different opioid manufacturers}
		\label{fig:market_share}
	\end{figure}

	\begin{figure} [H]
		\centering
	
		\begin{minipage}{0.8\textwidth}\
        \includegraphics[width=\linewidth]{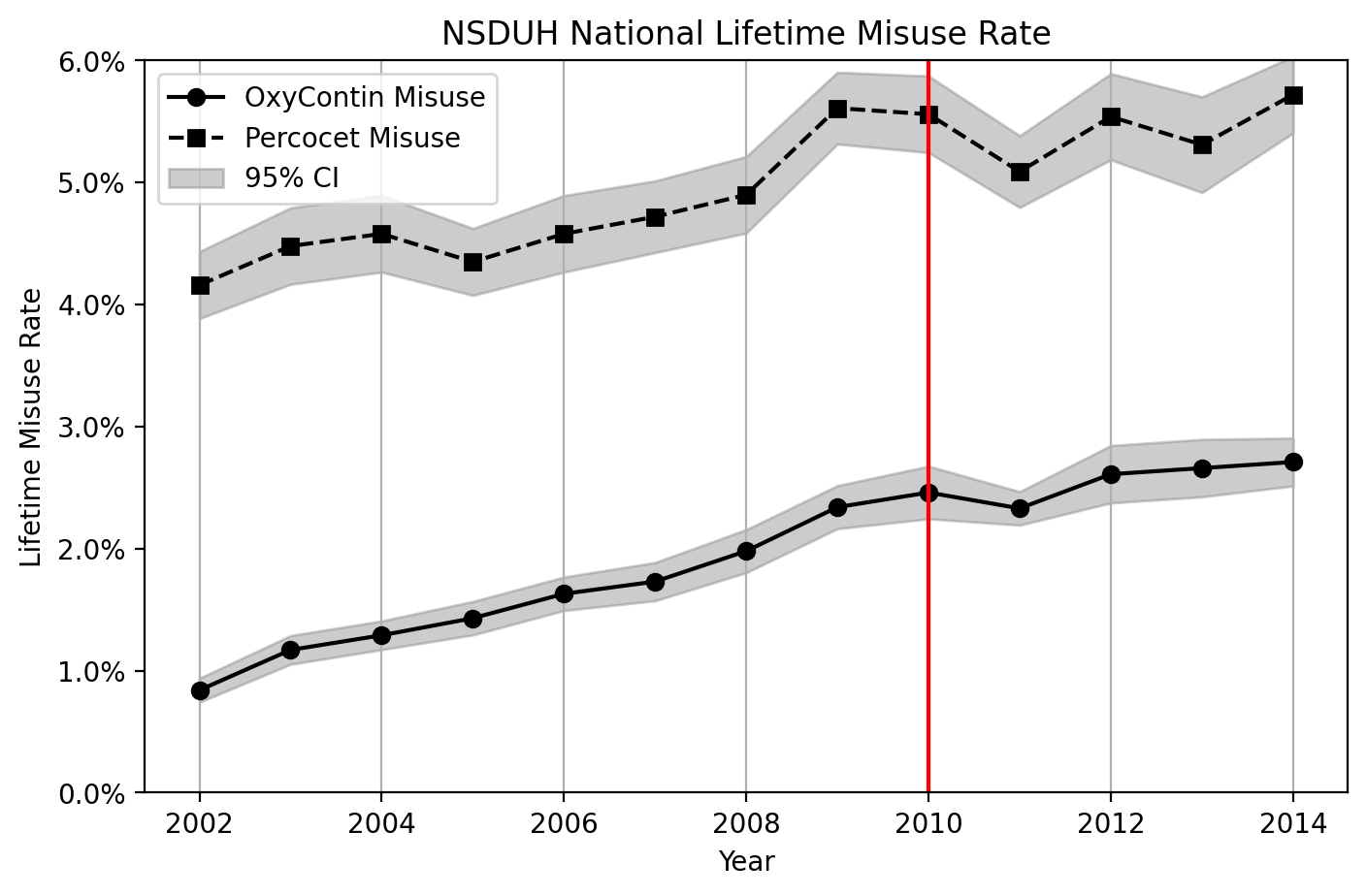}
        {\footnotesize \textit{Note:} The figure shows the misuse rate of OxyContin (OXYFLAG or OXYCONT2) and the misuse rate of Percocet, Percodan and Tylox (PERCTYL2). Data obtained from annual NSDUH. Percocet was a popular prescription oxycodone to misuse in the pre-OxyContin period. We see in this graph that the PERCTYL2 misuse rate increased 30\% from 2002 to 2009, suggesting that the lifetime misuse rate captures more than historical Percocet, Percodan and Tylox misuse. \par}
        \end{minipage}
		
		\caption{NSDUH national lifetime misuse rate}
		\label{fig:national_misuse}
	\end{figure}

	\begin{figure} [H]
		\centering
	
		\begin{minipage}{0.9\textwidth}\
        \includegraphics[width=\linewidth]{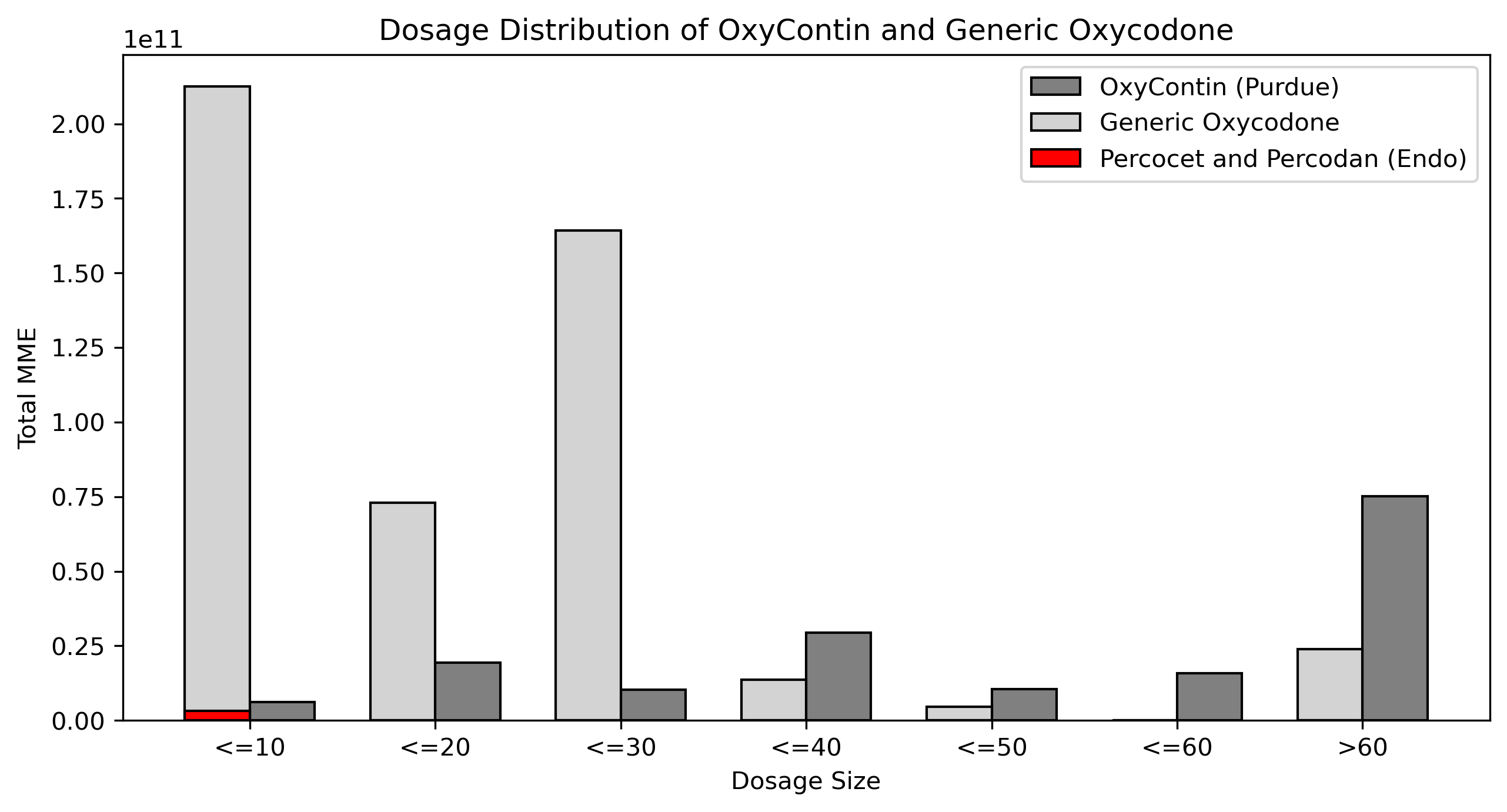}
        {\footnotesize \textit{Note:} This graph shows the difference in oxycodone sales between Purdue and Endo Pharma. The small market share of Endo Pharma leads us to believe that individuals misreport the drugs they consume on the NSDUH. \par}
        \end{minipage}
		
		\caption{Comparison of Sales of Purdue and Endo Pharma}
		\label{fig:purdue_endo}
	\end{figure}

	\begin{figure} [H]
		\centering
	
		\begin{minipage}{0.9\textwidth}\
        \includegraphics[width=\linewidth]{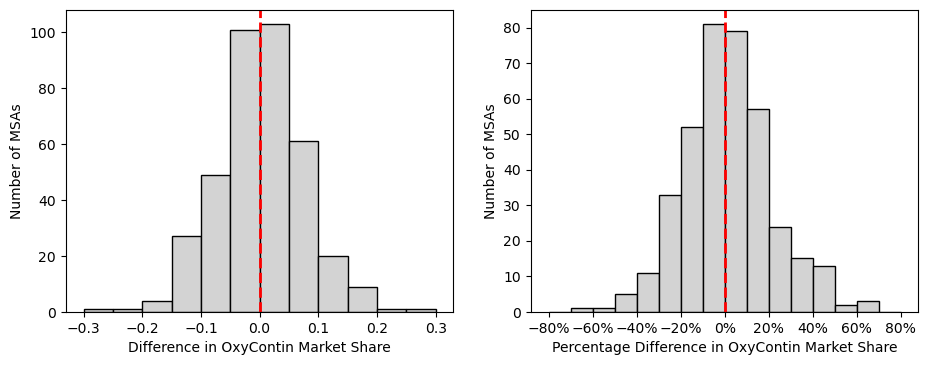}
        {\footnotesize \textit{Note:} Left is the absolute difference in market share (0.1 means that MSA share is 10\% higher than the state average) and right is percentage difference (10\% means that MSA share is 1.1 times the state average). \par}
        \end{minipage}
		
		\caption{Within-state variation in OxyContin Market Share}
		\label{fig:variation_oxy}
	\end{figure}

	\begin{figure} [H]
		\centering
	
		\begin{minipage}{0.9\textwidth}\
        \includegraphics[width=\linewidth]{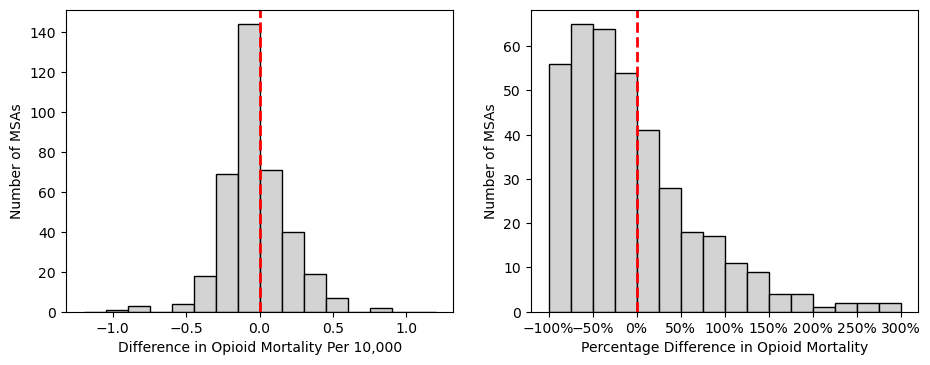}
        {\footnotesize \textit{Note:} Left is the absolute difference in opioid mortality (0.1 means that MSA mortality per 10,000 people is 0.1 higher than the state average) and right is percentage difference (10\% means that MSA mortality per 10,000 people is 1.1 times the state average) \par}
        \end{minipage}
		
		\caption{Within-state variation in opioid mortality}
		\label{fig:variation_mortality}
	\end{figure}
    
    \begin{figure} [H]
		\centering
	
		\begin{minipage}{0.9\textwidth}\
        \includegraphics[width=\linewidth]{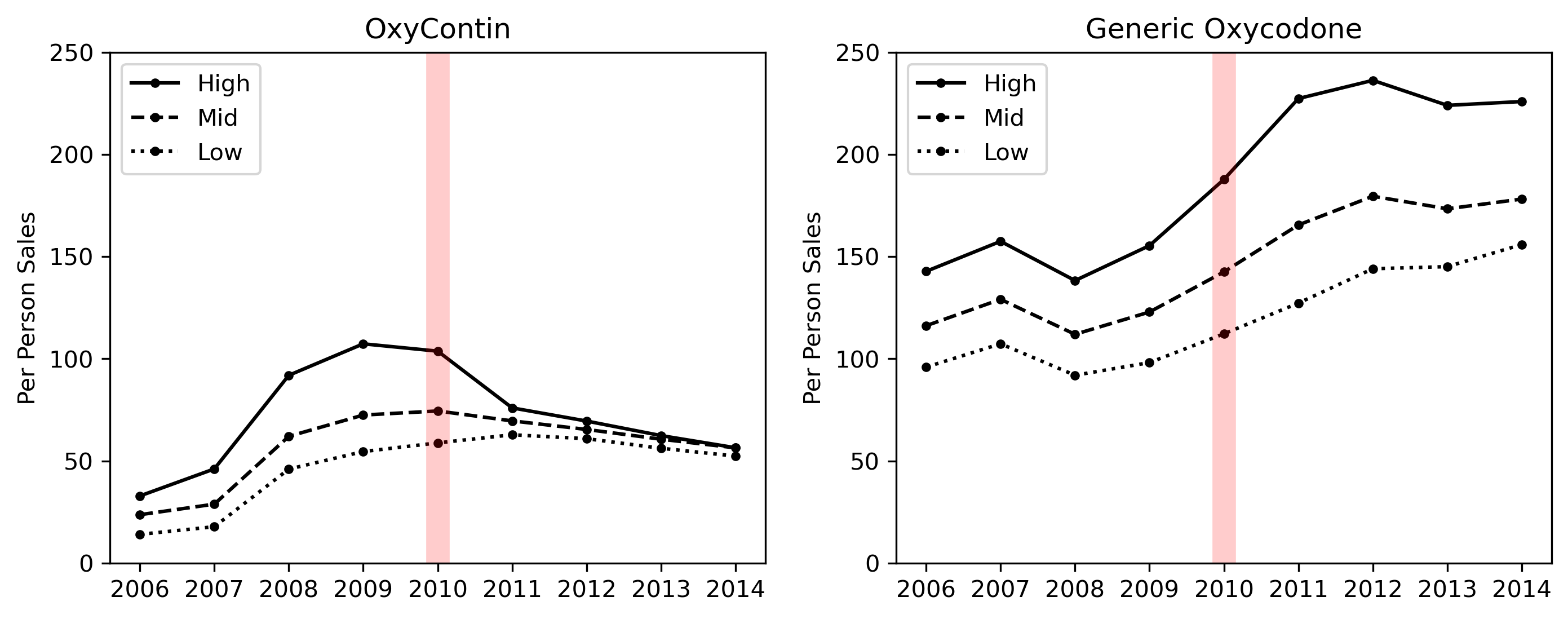}
        {\footnotesize \textit{Note:} We categorized all MSAs into high, mid, and low by the drop in the observed per person OxyContin sales from 2009 to 2011. The series are population weighted and Florida is excluded. The high group saw a 30\% drop in OxyContin sales, mid group a 3.9\% drop, and low group a 15\% increase. The high group experienced a 46\% increase in generic oxycodone sales, mid group a 34\% increase, and low group a 29\% increase. The three groups share similar oxycodone growth trends until the reformulation. 
         \par}
        \end{minipage}
		
		\caption{Opioid sales by empirical OxyContin drop}
		\label{fig:drop_1}
	\end{figure}
	
	\begin{figure} [H]
		\centering
	
		\begin{minipage}{0.9\textwidth}\
        \includegraphics[width=\linewidth]{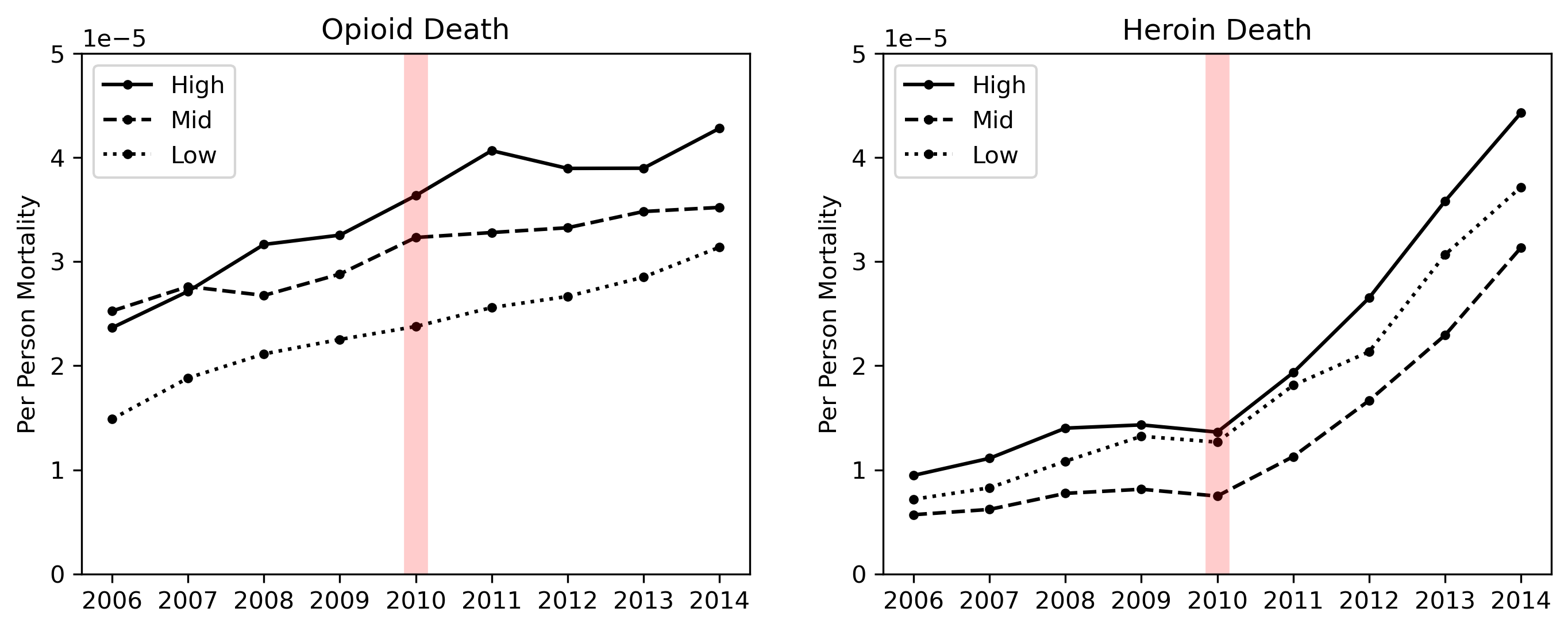}
        {\footnotesize \textit{Note:} Similarly to the previous figure, we categorized all MSAs into high, mid, and low by the drop in the observed per person OxyContin sales from 2009 to 2011. The series are population weighted and Florida is excluded. No trend break in opioid mortality in the high drop group. The high group saw an 35\% increase in heroin mortality, the mid group 38\%, and the low group 37\%. The similar increases in heroin mortality post-reform indicates that drops in OxyContin use post-reform did not lead to additional increase in heroin use.
         \par}
        \end{minipage}
		
		\caption{Opioid mortality by empirical OxyContin drop}
		\label{fig:drop_2}
	\end{figure}

\subsection{Tables}
    \begin{table}
        \begin{threeparttable}
        \footnotesize
        \renewcommand\arraystretch{1.1}
        \input{table1_justify_misuse}
        \end{threeparttable}
    \end{table}
    
    \begin{table}
        \begin{threeparttable}
        \footnotesize
        \renewcommand\arraystretch{1.1}
        \input{table3_DID_SE}
        \end{threeparttable}
    \end{table}

\subsection{Map}
	\begin{figure} [H]
		\centering
	
		\begin{minipage}{\textwidth}\
        \includegraphics[width=\linewidth]{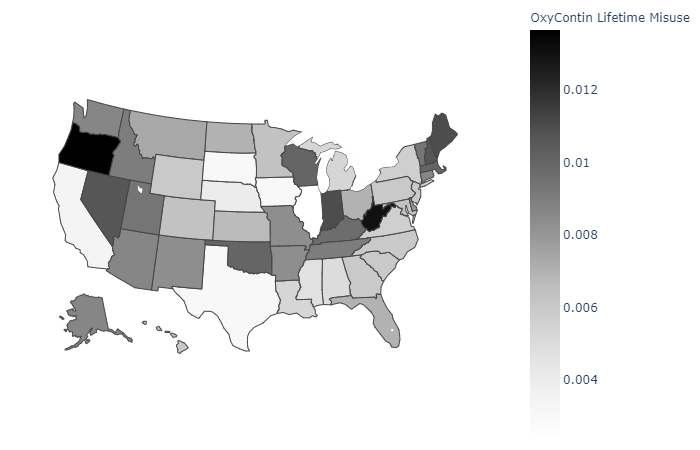}
        {\footnotesize \textit{Note:} Data from 2004-2009 NSDUH lifetime OxyContin misuse rate (NSDUH ticker OXXYR). 0.01 is interpreted as 1\% of the state population have ever misused OxyContin. \par}
        \end{minipage}
		
		\caption{OxyContin lifetime misuse rate at state level}
		\label{fig:map1}
	\end{figure}

	\begin{figure} [H]
		\centering
	
		\begin{minipage}{\textwidth}\
        \includegraphics[width=\linewidth]{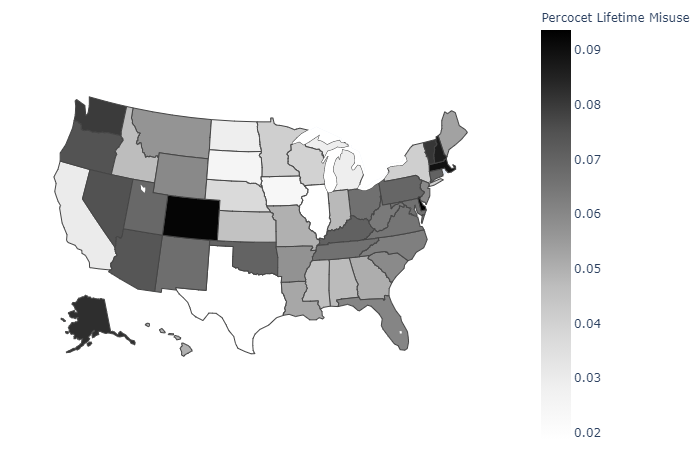}
        {\footnotesize \textit{Note:} Data from 2004-2009 NSDUH lifetime Percocet, Percodan, Tylox misuse rate (NSDUH ticker PERCTYL2). 0.01 is interpreted as 1\% of the state population have ever misused one of the three drugs. Percocet lifetime misuse rate on average is much higher than OxyContin lifetime misuse rate. \par}
        \end{minipage}
		
		\caption{Percocet lifetime misuse rate at state level}
		\label{fig:map2}
	\end{figure}

	\begin{figure} [H]
		\centering
	
		\begin{minipage}{\textwidth}\
        \includegraphics[width=\linewidth]{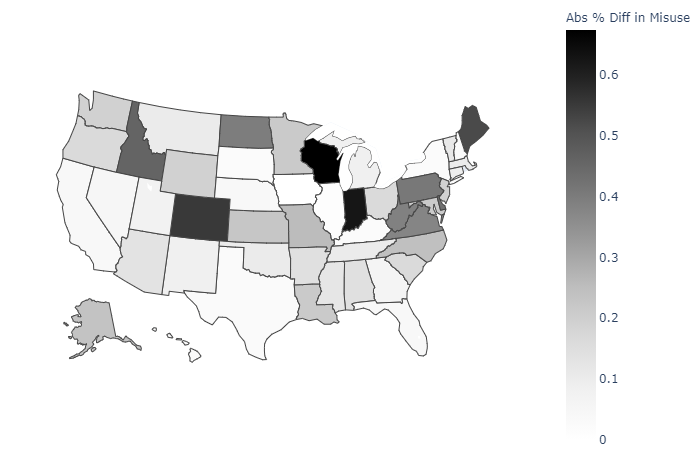}
        {\footnotesize \textit{Note:} The figure plots the absolute difference in percentile ranking of the two state level lifetime misuse rate. A 0.1 should be interpreted as a 10\% difference in percentile ranking between OxyContin lifetime misuse rate and Percocet lifetime misuse rate. For example, Colorado's OxyContin misuse rate is 0.0063  (42 percentile) and it's Percocet misuse rate is 0.092  (97 percentile), which is a 55\% difference in percentile ranking. We rely on the difference between two misuse rate to separately identify the impact of OxyContin and oxycodone. \par}
        \end{minipage}
		
		\caption{Difference in state level misuse rates}
		\label{fig:map3}
	\end{figure}
	
	\begin{figure} [H]
		\centering
	
		\begin{minipage}{\textwidth}\
        \includegraphics[width=\linewidth]{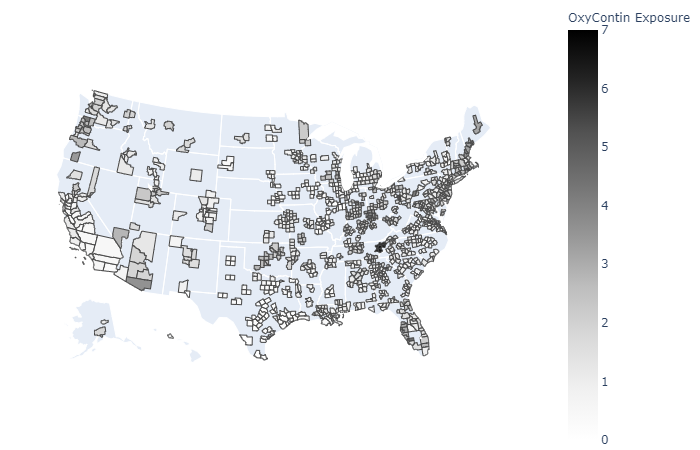}
        {\footnotesize \textit{Note:} This figure shows OxyContin exposure by MSA. We show Florida here, which had very low OxyContin exposure/sales, but omit it from analysis because it had abnormally high generic oxycodone sales with large amounts being trafficked to other states. \par}
        \end{minipage}
		
		\caption{OxyContin exposure at MSA level}
		\label{fig:map4}
	\end{figure}

	\begin{figure} [H]
		\centering
	
		\begin{minipage}{\textwidth}\
        \includegraphics[width=\linewidth]{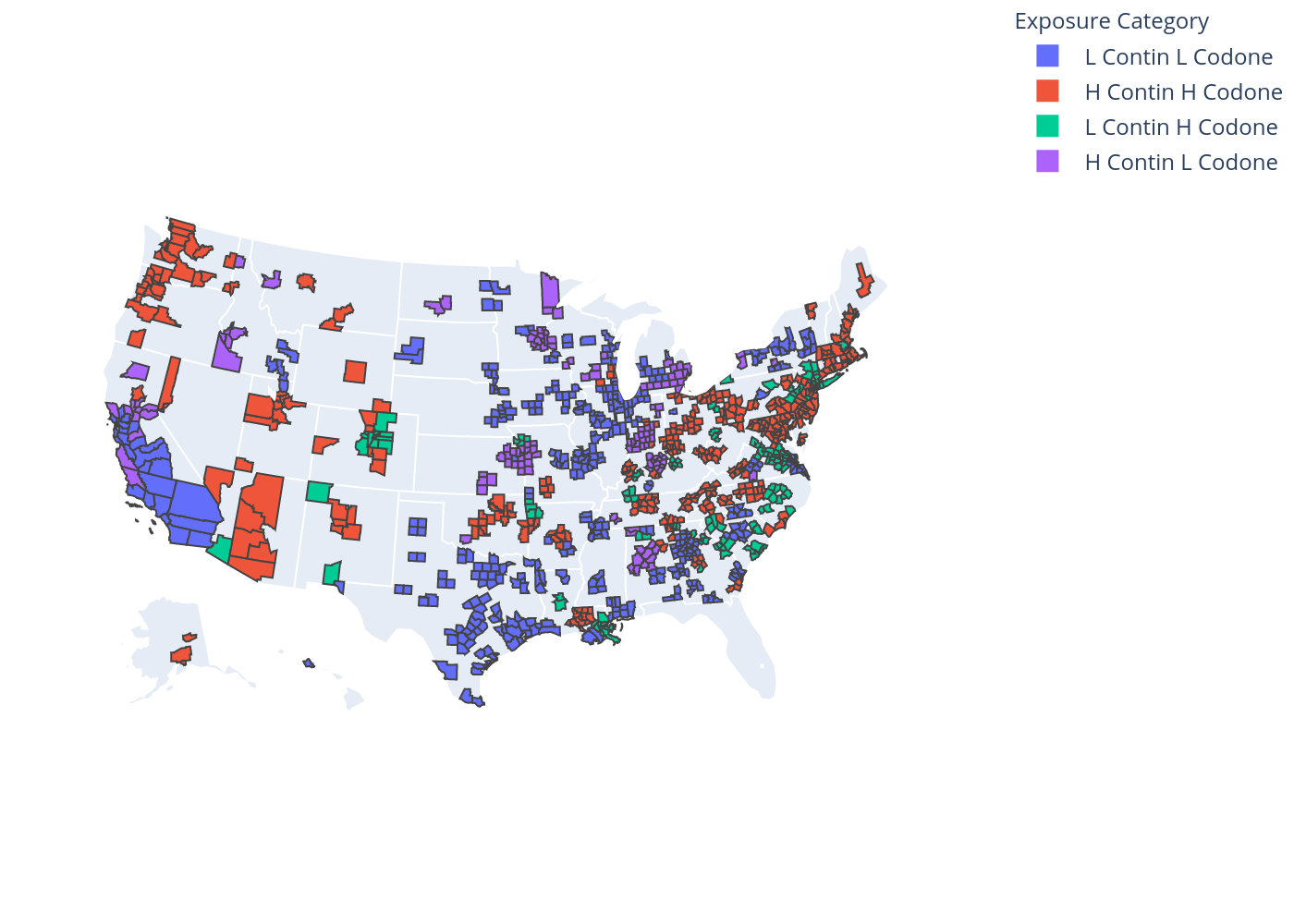}
        {\footnotesize \textit{Note:} Florida is excluded in this analysis. MSAs grouped by high vs low OxyContin exposure and high vs low generic oxycodone exposure.\par}
        \end{minipage}
		
		\caption{Diff-in-diff regression categories}
		\label{fig:map44}
	\end{figure}

\subsection{Alternative Regression Specifications}
\subsubsection{MSA FE}

	\begin{figure} [H]
		\centering
	
		\begin{minipage}{0.8\textwidth}\
        \includegraphics[width=\linewidth]{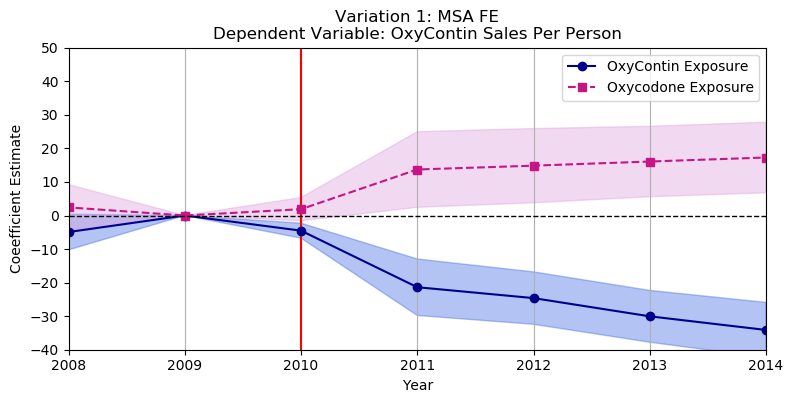}
        \end{minipage}
		
		\caption{Regression on OxyContin sales with MSA FE. Shaded regions are the 95 percent confidence intervals with standard errors clustered at the MSA level.}
	\end{figure}
	
	\begin{figure} [H]
		\centering
	
		\begin{minipage}{0.8\textwidth}\
        \includegraphics[width=\linewidth]{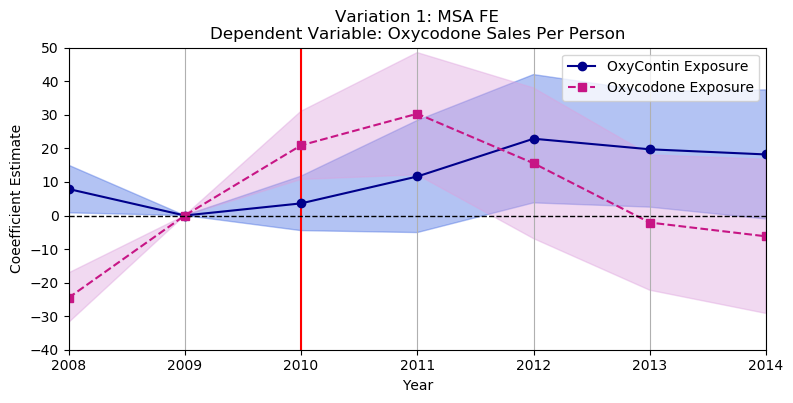}
        \end{minipage}
		
		\caption{Regression on oxycodone sales with MSA FE. Shaded regions are the 95 percent confidence intervals with standard errors clustered at the MSA level.}
	\end{figure}
	
	\begin{figure} [H]
		\centering
	
		\begin{minipage}{0.8\textwidth}\
        \includegraphics[width=\linewidth]{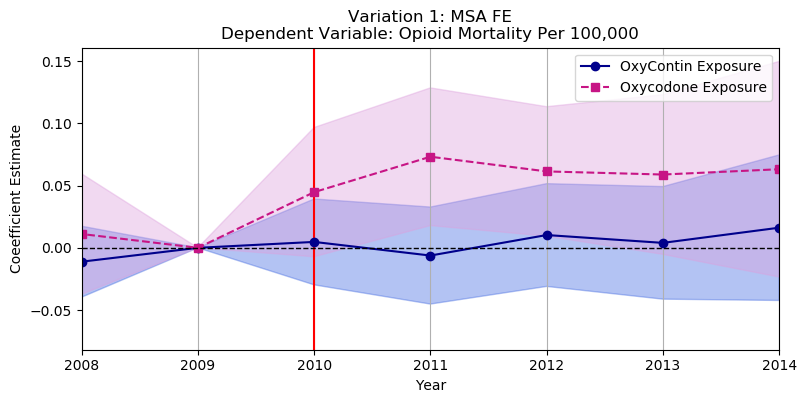}
        \end{minipage}
		
		\caption{Regression on opioid mortality with MSA FE. Shaded regions are the 95 percent confidence intervals with standard errors clustered at the MSA level.}
	\end{figure}

	\begin{figure} [H]
		\centering
	
		\begin{minipage}{0.8\textwidth}\
        \includegraphics[width=\linewidth]{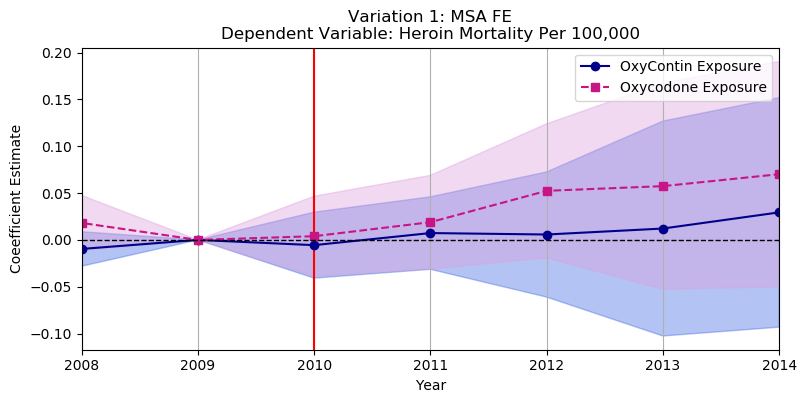}
        \end{minipage}
		
		\caption{Regression on heroin mortality with MSA FE. Shaded regions are the 95 percent confidence intervals with standard errors clustered at the MSA level.}
	\end{figure}
	
\subsubsection{Last Year OxyContin Misuse}
\label{sec:OXYYR}

	\begin{figure} [H]
		\centering
	
		\begin{minipage}{0.8\textwidth}\
        \includegraphics[width=\linewidth]{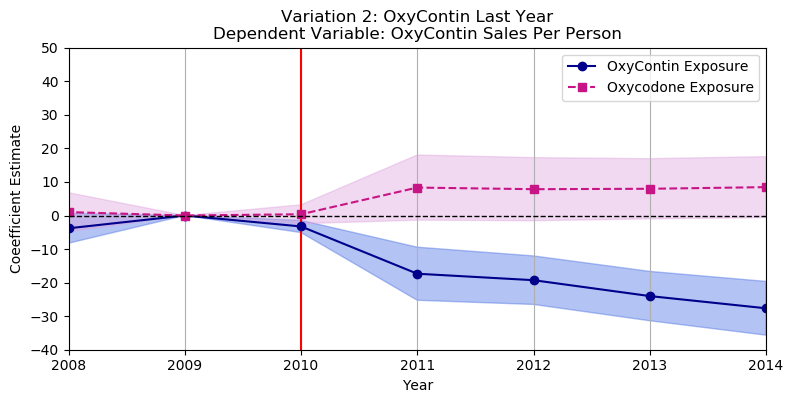}
        \end{minipage}
		
		\caption{Regression on OxyContin sales with last-year OxyContin. Shaded regions are the 95 percent confidence intervals with standard errors clustered at the MSA level.}
	\end{figure}
	
	\begin{figure} [H]
		\centering
	
		\begin{minipage}{0.8\textwidth}\
        \includegraphics[width=\linewidth]{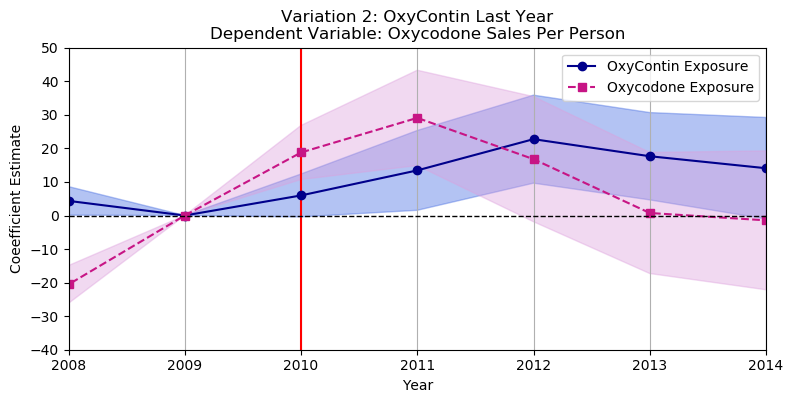}
        \end{minipage}
		
		\caption{Regression on oxycodone sales with last-year OxyContin. Shaded regions are the 95 percent confidence intervals with standard errors clustered at the MSA level.}
	\end{figure}
	
	\begin{figure} [H]
		\centering
	
		\begin{minipage}{0.8\textwidth}\
        \includegraphics[width=\linewidth]{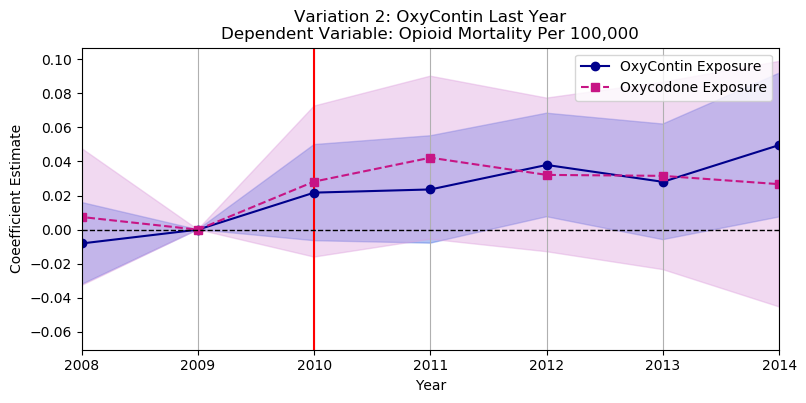}
        \end{minipage}
		
		\caption{Regression on opioid mortality with last-year OxyContin. Shaded regions are the 95 percent confidence intervals with standard errors clustered at the MSA level.}
	\end{figure}

	\begin{figure} [H]
		\centering
	
		\begin{minipage}{0.8\textwidth}\
        \includegraphics[width=\linewidth]{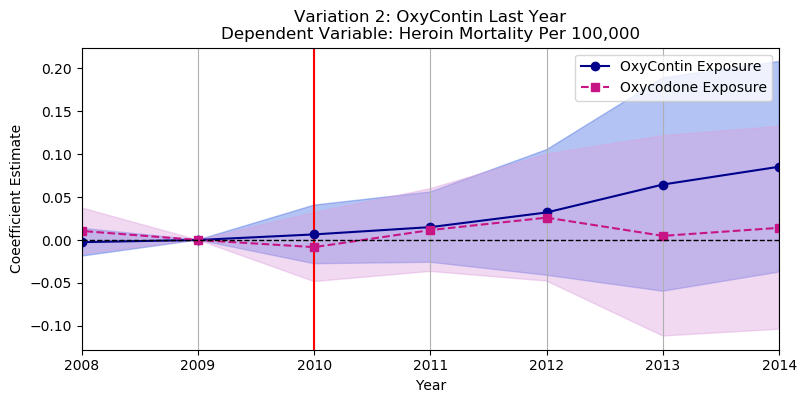}
        \end{minipage}
		
		\caption{Regression on heroin mortality with last-year OxyContin. Shaded regions are the 95 percent confidence intervals with standard errors clustered at the MSA level.}
	\end{figure}

\subsubsection{State Level Regression}

\label{sec:state}
	\begin{figure} [H]
		\centering
	
		\begin{minipage}{0.8\textwidth}\
        \includegraphics[width=\linewidth]{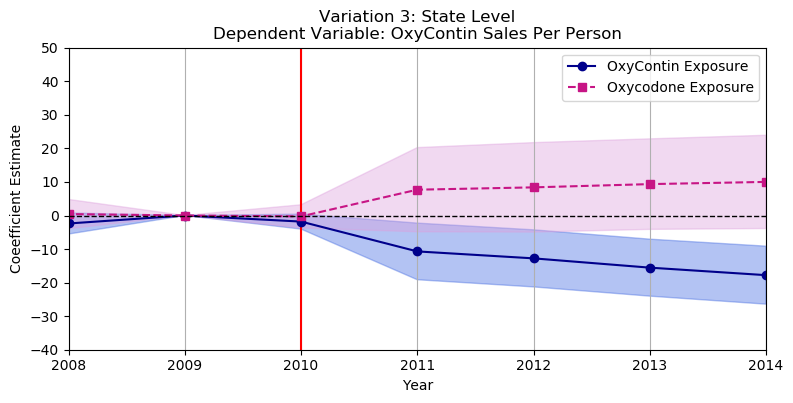}
        \end{minipage}
    		
		\caption{Regression on OxyContin sales at state level. Shaded regions are the 95 percent confidence intervals with standard errors clustered at the MSA level.}
	\end{figure}
	
	\begin{figure} [H]
		\centering
	
		\begin{minipage}{0.8\textwidth}\
        \includegraphics[width=\linewidth]{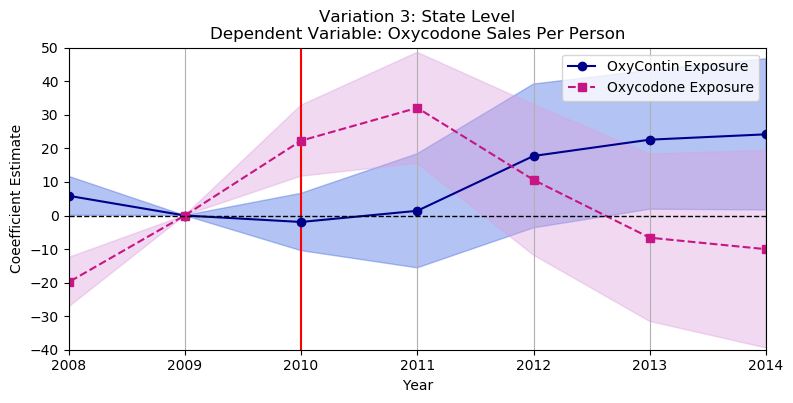}
        \end{minipage}
		
		\caption{Regression on oxycodone sales at state level. Shaded regions are the 95 percent confidence intervals with standard errors clustered at the MSA level.}
	\end{figure}
	
	\begin{figure} [H]
		\centering
	
		\begin{minipage}{0.8\textwidth}\
        \includegraphics[width=\linewidth]{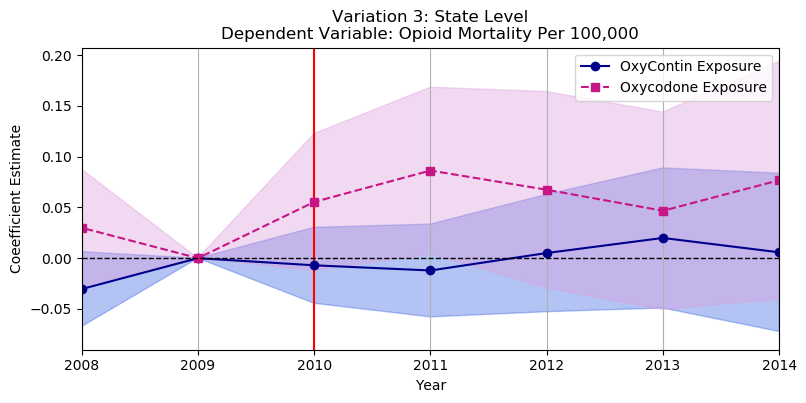}
        \end{minipage}
		
		\caption{Regression on opioid mortality at state level}
	\end{figure}

	\begin{figure} [H]
		\centering
	
		\begin{minipage}{0.8\textwidth}\
        \includegraphics[width=\linewidth]{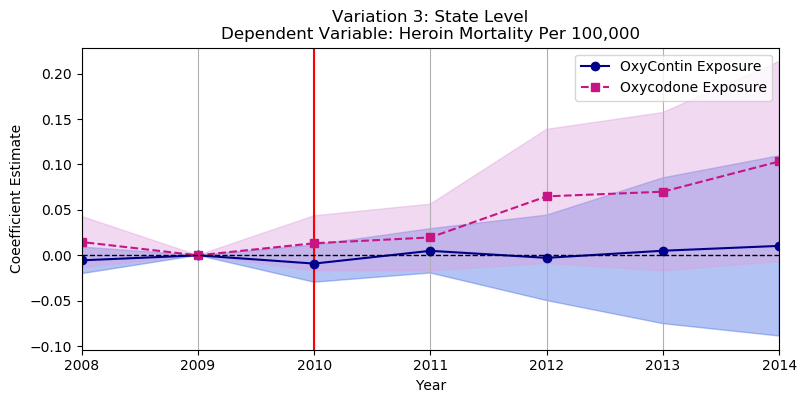}
        \end{minipage}
		
		\caption{Regression on heroin mortality at state level. Shaded regions are the 95 percent confidence intervals with standard errors clustered at the MSA level.}
	\end{figure}
	
\subsubsection{OxyContin Only}
\label{sec:only_oxy}

	\begin{figure} [H]
		\centering
	
		\begin{minipage}{0.8\textwidth}\
        \includegraphics[width=\linewidth]{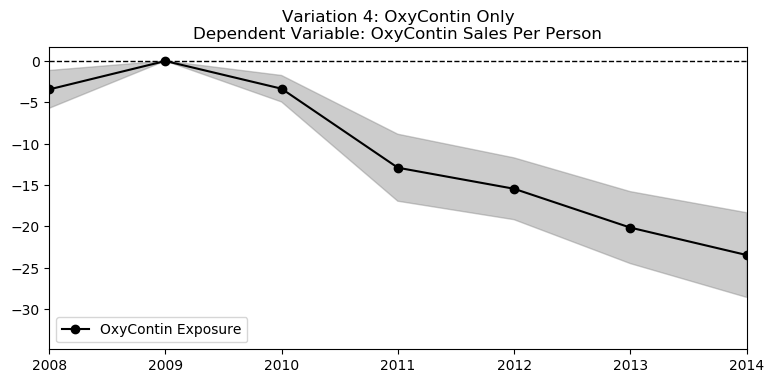}
        \end{minipage}
		
		\caption{Regression on OxyContin sales with OxyContin only. Shaded regions are the 95 percent confidence intervals with standard errors clustered at the MSA level.}
	\end{figure}
	
	\begin{figure} [H]
		\centering
	
		\begin{minipage}{0.8\textwidth}\
        \includegraphics[width=\linewidth]{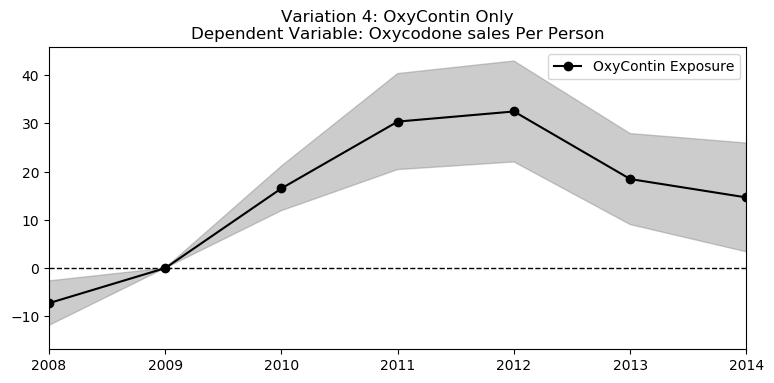}
        \end{minipage}
		
		\caption{Regression on oxycodone sales with OxyContin only. Shaded regions are the 95 percent confidence intervals with standard errors clustered at the MSA level.}
	\end{figure}
	
	\begin{figure} [H]
		\centering
	
		\begin{minipage}{0.8\textwidth}\
        \includegraphics[width=\linewidth]{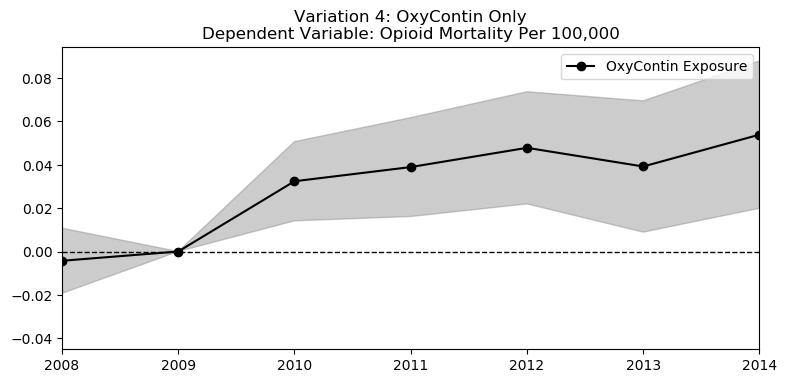}
        \end{minipage}
		
		\caption{Regression on opioid mortality with OxyContin only. Shaded regions are the 95 percent confidence intervals with standard errors clustered at the MSA level.}
	\end{figure}

	\begin{figure} [H]
	
		\centering
	
		\begin{minipage}{0.8\textwidth}\
        \includegraphics[width=\linewidth]{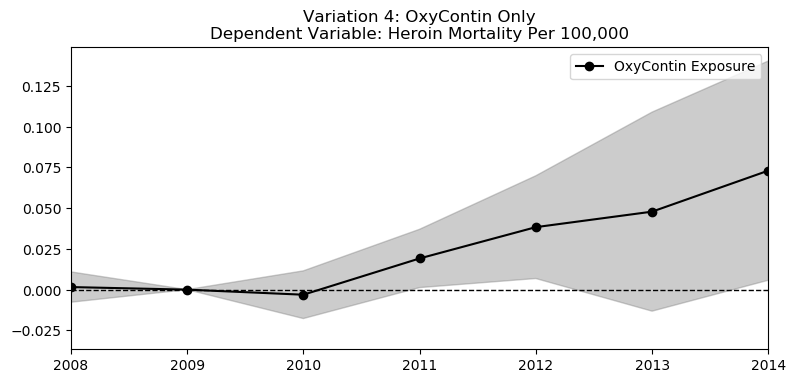}
        \end{minipage}
		
		\caption{Regression on heroin mortality with OxyContin only. Shaded regions are the 95 percent confidence intervals with standard errors clustered at the MSA level.}
		\label{single_heroin1}
	\end{figure}

\subsubsection{Oxycodone Only}
\label{sec:only_codone}
	\begin{figure} [H]
		\centering
	
		\begin{minipage}{0.8\textwidth}\
        \includegraphics[width=\linewidth]{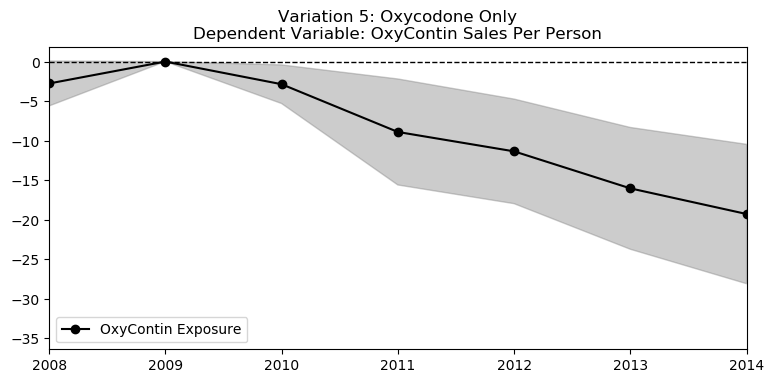}
        \end{minipage}
		
		\caption{Regression on OxyContin sales with oxycodone only. Shaded regions are the 95 percent confidence intervals with standard errors clustered at the MSA level.}
	\end{figure}
	
	\begin{figure} [H]
		\centering
	
		\begin{minipage}{0.8\textwidth}\
        \includegraphics[width=\linewidth]{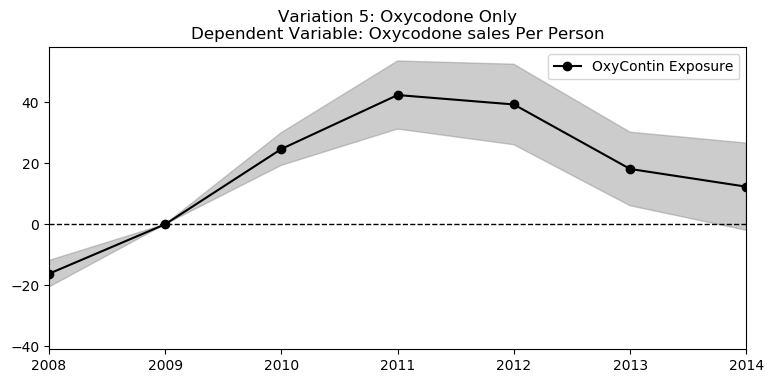}
        \end{minipage}
		
		\caption{Regression on oxycodone sales with oxycodone only. Shaded regions are the 95 percent confidence intervals with standard errors clustered at the MSA level.}
	\end{figure}
	
	\begin{figure} [H]
		\centering
	
		\begin{minipage}{0.8\textwidth}\
        \includegraphics[width=\linewidth]{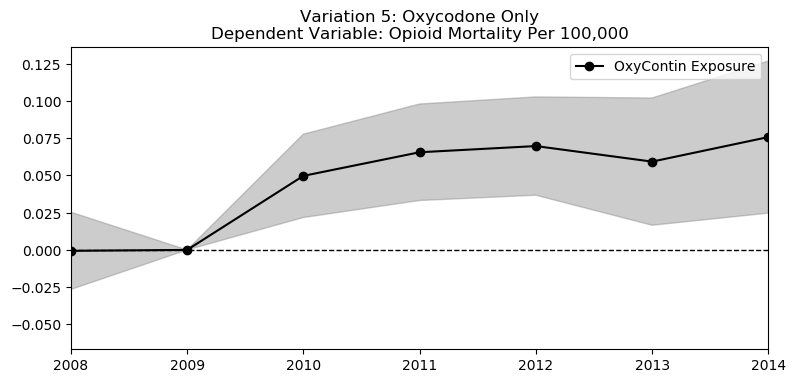}
        \end{minipage}
		
		\caption{Regression on opioid mortality with oxycodone only. Shaded regions are the 95 percent confidence intervals with standard errors clustered at the MSA level.}
	\end{figure}

	\begin{figure} [H]
	
		\centering
	
		\begin{minipage}{0.8\textwidth}\
        \includegraphics[width=\linewidth]{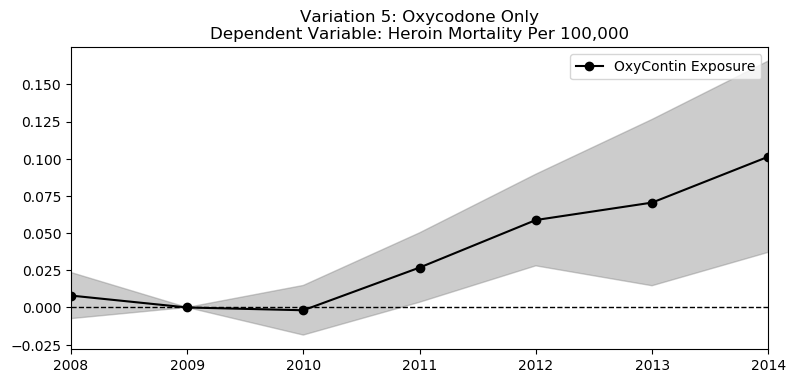}
        \end{minipage}
		
		\caption{Regression on heroin mortality with oxycodone only. Shaded regions are the 95 percent confidence intervals with standard errors clustered at the MSA level.}
		\label{single_heroin2}
	\end{figure}	

\end{document}

%% file: table2_summary_stat.tex
\begin{center}
\caption{Summary Statistics}
\label{tab:table2}
\begin{tabular} {L{0.01\textwidth}L{0.3\textwidth}C{0.11\textwidth}C{0.11\textwidth}C{0.11\textwidth}C{0.11\textwidth}C{0.11\textwidth}}

\hline\hline
&& All MSAs & MSAs with low OxyContin exposure & MSAs with high OxyContin exposure & MSAs with low oxycodone exposure & MSAs with high oxycodone exposure \\ \hline
\multicolumn{7}{l}{\textit{NSDUH lifetime misuse rates (2004-2009)}} \\
& OxyContin misuse rate (\%) & 2.22 & 1.88 & 2.56 & 1.87 & 2.56 \\
& Oxycodone misuse rate (\%) & 5.19 & 4.22 & 6.17 & 3.75 & 6.64 \\[0.1cm]
\multicolumn{7}{l}{\textit{Annual ARCOS sales (all sample period)}} \\
& Oxycontin sales per person & 65.71 & 43.47 & 88.06 & 50.70 & 80.79 \\
& Oxycodone sales per person & 181.84 & 112.50 & 251.55 & 99.24 & 264.88 \\[0.1cm]
\multicolumn{7}{l}{\textit{Annual death per 100,000 (all sample period)}} \\
& Opioid & 0.32 & 0.23 & 0.41 & 0.23 & 0.42 \\
& Heroin & 0.13 & 0.09 & 0.16 & 0.10 & 0.16 \\[0.1cm]
\multicolumn{7}{l}{\textit{Census Demographics (2009)}} \\
& Number of MSAs & 379 & 190 & 189 & 190 & 189 \\
& Population & 679878 & 745327 & 614082 & 663740 & 696101 \\
& Age & 36.13 & 34.68 & 37.59 & 34.84 & 37.43 \\
& Male (\%) & 49.24 & 49.35 & 49.13 & 49.40 & 49.08 \\
& Separated (\%) & 18.83 & 18.24 & 19.42 & 18.32 & 19.34 \\
& High school and above (\%) & 84.20 & 82.79 & 85.61 & 83.68 & 84.72 \\
& Bachelor and above (\%) & 25.36 & 24.77 & 25.96 & 24.85 & 25.87 \\
& Mean income & 64213 & 63414 & 65016 & 63058 & 65374 \\
& Low income (\%) & 35.38 & 35.79 & 34.98 & 35.90 & 34.86 \\
& White (\%) & 82.17 & 79.99 & 84.36 & 81.22 & 83.12 \\
& Black (\%) & 11.20 & 13.09 & 9.30 & 11.80 & 10.60 \\
& Asian (\%) & 3.03 & 3.47 & 2.60 & 3.52 & 2.54 \\
& Native American (\%) & 0.18 & 0.20 & 0.17 & 0.20 & 0.17 \\\hline
\end{tabular}

\begin{tablenotes}
      \small
      \item \textit{Note:} Simple average, not weighted by population.
\end{tablenotes}

\end{center}

%% file: table1_justify_misuse.tex
\begin{center}
\caption{Testing Constructed Exposure Measure Against Opioid Mortality}
\label{tab:justification}
\begin{tabular} {L{0.3\textwidth}C{0.08\textwidth}C{0.08\textwidth}C{0.08\textwidth}C{0.08\textwidth}C{0.08\textwidth}C{0.08\textwidth}}
\hline\hline
 & \multicolumn{6}{c}{Opioid overdose deaths per 100,000}\\\cline{2-7}
 &\multicolumn{3}{c}{OxyContin} & \multicolumn{3}{c}{Generic Oxycodone}\\
 & (1) & (2) & (3) & (4) & (5) & (6)\\\hline
 NSDUH misuse & 10.235 & & & 2.909 & &\\
 & (1.719) & & & (0.570) & &\\
 ARCOS sales & & 0.001 & & & 0.001 &\\
 & & (0.0002) & & & (0.0001) &\\
 Combined exposure & & & 0.093 & & & 0.087\\
 & & & (0.012) & & & (0.009)\\
 Number of observations & 379 & 379 & 379 & 379 & 379 & 379\\
 R-square & 0.086 & 0.089 & 0.130 & 0.065 & 0.178 & 0.189\\
 Adjusted R-square & 0.084 & 0.086 & 0.128 & 0.062 & 0.176 & 0.187 \\\hline
\end{tabular}

\begin{tablenotes}
      \small
      \item \textit{Notes:} Standard errors are in parentheses. We report coefficients from OLS regressions of opioid mortality on misuse, sales or exposure. NSDUH misuse rates is the 6-year average OxyContin or Percocet lifetime misuse rate from pre-reform period (2004-2009). ACROS sales is Oxycontin or generic oxycodone sales per person from 2009. Combined exposure is the product of the previous two measures normalized (see equation 1). Overdose from 2009. Regressions are weighted by MSA population. 
\end{tablenotes}

\end{center}

%% file: table3_DID_SE.tex
\begin{center}
\caption{Difference in difference regression results}
\label{tab:did}
\begin{tabular} {L{0.3\textwidth}C{0.13\textwidth}C{0.13\textwidth}C{0.13\textwidth}C{0.13\textwidth}}
\hline\hline
 & \multicolumn{2}{c}{Opioid sales per person} & \multicolumn{2}{c}{Overdose per 10,000} \\ \cline{2-5}
 & OxyContin & Oxycodone & Opioid & Heroin \\
 & (1) & (2) & (3) & (4) \\ \hline
Post & -8.05 & 41.74 & 0.01 & 0.14 \\
 & (2.86) & (4.92) & (0.02) & (0.02) \\
High OxyContin & 47.24 & 56.46 & -0.05 & -0.07 \\
 & (5.78) & (13.36) & (0.03) & (0.02) \\
High Oxycodone & 26.84 & 95.90 & 0.14 & 0.08 \\
 & (6.66) & (15.47) & (0.04) & (0.05) \\
Post x High OxyContin & -15.14 & 10.30 & 0.02 & 0.03 \\
 & (6.39) & (8.90) & (0.02) & (0.02) \\
Post x High Oxycodone & -2.33 & 33.99 & 0.06 & 0.07 \\
 & (6.37) & (8.80) & (0.02) & (0.02) \\
Number of observations & 2148 & 2148 & 2148 & 2148 \\
R-square & 0.665 &  0.737 & 0.517 & 0.469 \\
Adjusted R-square & 0.654 & 0.728 & 0.501 & 0.452 \\\hline
\end{tabular}

\begin{tablenotes}
      \small
      \item \textit{Notes:} We report coefficients from the difference-in-difference estimation (see equation (3)). All MSAs in Florida are excluded. In all specifications, we include MSA-level control variables, state fixed effects and year fixed effects. Standard errors are clustered at the MSA level.
\end{tablenotes}

\end{center}

%% file: ref.bib
@article{Cicero2015,
    author = {Cicero, Theodore and Ellis, Matthew},
    title = "{Abuse-Deterrent Formulations and the Prescription Opioid Abuse Epidemic in the United States: Lessons Learned From OxyContin}",
    journal = {JAMA Psychiatry},
    volume = {72},
    number = {5},
    pages = {424-430},
    year = {2015},
    month = {05},
    issn = {2168-622X},
    doi = {10.1001/jamapsychiatry.2014.3043},
    url = {https://doi.org/10.1001/jamapsychiatry.2014.3043},
    eprint = {https://jamanetwork.com/journals/jamapsychiatry/articlepdf/2174541/yoi140121.pdf},
}

@article{Alpert2017,
 title = "Supply-Side Drug Policy in the Presence of Substitutes:  Evidence from the Introduction of Abuse-Deterrent Opioids",
 author = "Alpert, Abby and Powell, David and Pacula, Rosalie Liccardo",
 institution = "National Bureau of Economic Research",
 type = "Working Paper",
 series = "Working Paper Series",
 number = "23031",
 year = "2017",
 month = "January",
 doi = {10.3386/w23031},
 URL = "http://www.nber.org/papers/w23031",
}

@article{alpert2019origin,
 title = "Origins of the Opioid Crisis and Its Enduring Impacts",
 %author = "Alpert, Abby and Evans, William and Lieber, Ethan and Powell, David",
 author = "Alpert, Abby and Evans and Lieber and Powell",
 institution = "National Bureau of Economic Research",
 type = "Working Paper",
 series = "Working Paper Series",
 number = "26500",
 year = "2019",
 month = "11",
 doi = {10.3386/w26500},
 URL = "http://www.nber.org/papers/w26500",
}

@article{inciardi2009black,
  title={The “black box” of prescription drug diversion},
  author={Inciardi, James and Surratt, Hilary and Cicero, Theodore and Kurtz, Steven and Martin, Steven and Parrino, Mark},
  journal={Journal of addictive diseases},
  volume={28},
  number={4},
  pages={332--347},
  year={2009},
  publisher={Taylor \& Francis}
}

@article{Schnell2017,
  title={Physician behavior in the presence of a secondary market: The case of prescription opioids},
  author={Schnell, Molly},
  year = "2017"
}

@article{hays2004profile,
  title={A profile of OxyContin addiction},
  author={Hays, Lon R},
  journal={Journal of Addictive Diseases},
  volume={23},
  number={4},
  pages={1--9},
  year={2004},
  publisher={Taylor \& Francis}
}

@article{sproule2009changing,
  title={Changing patterns in opioid addiction: characterizing users of oxycodone and other opioids},
  author={Sproule, Beth and Brands, Bruna and Li, Selina and Catz-Biro, Laura},
  journal={Canadian Family Physician},
  volume={55},
  number={1},
  pages={68--69},
  year={2009},
  publisher={The College of Family Physicians of Canada}
}

@article{alpert2018supply,
  title={Supply-side drug policy in the presence of substitutes: Evidence from the introduction of abuse-deterrent opioids},
  author={Alpert, Abby and Powell, David and Pacula, Rosalie Liccardo},
  journal={American Economic Journal: Economic Policy},
  volume={10},
  number={4},
  pages={1--35},
  year={2018}
}

@article{mallatt2018effect,
  title={The effect of prescription drug monitoring programs on opioid prescriptions and heroin crime rates},
  author={Mallatt, Justine},
  journal={Available at SSRN 3050692},
  year={2018}
}

@article{sairam2014assessment,
  title={Assessment of the trends in medical use and misuse of opioid analgesics from 2004 to 2011},
  author={Atluri, Sairam and Sundarshan, G and Manchikanti, Laxmaiah},
  journal={ASIPP},
  volume={17},
  pages={E119--E28},
  year={2014}
}

@article{webster2009update,
  title={Update on abuse-resistant and abuse-deterrent approaches to opioid formulations},
  author={Webster, Lynn},
  journal={Pain Medicine},
  volume={10},
  number={suppl\_2},
  pages={S124--S133},
  year={2009},
  publisher={Blackwell Publishing Inc Malden, USA}
}

@report{oxymedreview2009,
title = {APPLICATION NUMBER:
22-272 - MEDICAL REVIEW(S)},
author = {Rappaport, Bob},
institution = {FDA CENTER FOR DRUG EVALUATION AND RESEARCH},
year = {2009},
url = {https://www.accessdata.fda.gov/drugsatfda_docs/nda/2010/022272s000MedR.pdf}
}

@report{nsduhMSA2012,
title = {2005-2010 NSDUH MSA Detailed Tables},
intsitution = {U.S. DEPARTMENT OF HEALTH AND HUMAN SERVICES
Substance Abuse and Mental Health Services Administration
Center for Behavioral Health Statistics and Quality},
year = {2012},
url = {https://web.archive.org/web/20130616122357/http://www.samhsa.gov/data/NSDUHMetroBriefReports/index.aspx}
}

@article{van2009promotion,
  title={The promotion and marketing of oxycontin: commercial triumph, public health tragedy},
  author={Van Zee, Art},
  journal={American journal of public health},
  volume={99},
  number={2},
  pages={221--227},
  year={2009},
  publisher={American Public Health Association}
}

@article{evans2019reformulation,
  title={How the reformulation of OxyContin ignited the heroin epidemic},
  author={Evans, William and Lieber, Ethan and Power, Patrick},
  journal={Review of Economics and Statistics},
  volume={101},
  number={1},
  pages={1--15},
  year={2019},
  publisher={MIT Press}
}

@article{cicero2012effect,
  title={Effect of abuse-deterrent formulation of OxyContin},
  author={Cicero, Theodore and Ellis, Matthew and Surratt, Hilary},
  journal={New England Journal of Medicine},
  volume={367},
  number={2},
  pages={187--189},
  year={2012},
  publisher={Mass Medical Soc}
}

@article{bloomquist1963addiction,
  title={The addiction potential of oxycodone (Percodan{\textregistered})},
  author={Bloomquist, Edward R},
  journal={California medicine},
  volume={99},
  number={2},
  pages={127},
  year={1963},
  publisher={BMJ Publishing Group}
}

@article{quinn1965percodan,
  title={Percodan on Triplicate—The Background of the New Law},
  author={Quinn, Wm F},
  journal={California medicine},
  volume={103},
  number={3},
  pages={212},
  year={1965},
  publisher={BMJ Publishing Group}
}

@article{mars2014every,
  title={“Every ‘never’I ever said came true”: transitions from opioid pills to heroin injecting},
  author={Mars, Sarah G and Bourgois, Philippe and Karandinos, George and Montero, Fernando and Ciccarone, Daniel},
  journal={International Journal of Drug Policy},
  volume={25},
  number={2},
  pages={257--266},
  year={2014},
  publisher={Elsevier}
}

@article{ruhm2017geographic,
  title={Geographic variation in opioid and heroin involved drug poisoning mortality rates},
  author={Ruhm, Christopher J},
  journal={American journal of preventive medicine},
  volume={53},
  number={6},
  pages={745--753},
  year={2017},
  publisher={Elsevier}
}

@article{modarai2013relationship,
  title={Relationship of opioid prescription sales and overdoses, North Carolina},
  author={Modarai, F and Mack, K and Hicks, P and Benoit, S and Park, S and Jones, C and Proescholdbell, S and Ising, A and Paulozzi, L},
  journal={Drug and alcohol dependence},
  volume={132},
  number={1-2},
  pages={81--86},
  year={2013},
  publisher={Elsevier}
}

@article{paulozzi2006opioid,
  title={Opioid analgesics and rates of fatal drug poisoning in the United States},
  author={Paulozzi, Leonard J and Ryan, George W},
  journal={American journal of preventive medicine},
  volume={31},
  number={6},
  pages={506--511},
  year={2006},
  publisher={Elsevier}
}

@article{coplan2013changes,
  title={Changes in oxycodone and heroin exposures in the National Poison Data System after introduction of extended-release oxycodone with abuse-deterrent characteristics},
  author={Coplan, Paul M and Kale, Hrishikesh and Sandstrom, Lauren and Landau, Craig and Chilcoat, Howard D},
  journal={Pharmacoepidemiology and drug safety},
  volume={22},
  number={12},
  pages={1274--1282},
  year={2013},
  publisher={Wiley Online Library}
}

@article{siegal2003probable,
  title={Probable relationship between opioid abuse and heroin use},
  author={Siegal, Harvey A and Carlson, Robert G and Kenne, Deric R and Swora, Maria G},
  journal={American family physician},
  volume={67},
  number={5},
  pages={942},
  year={2003}
}

@article{compton2016relationship,
  title={Relationship between nonmedical prescription-opioid use and heroin use},
  author={Compton, Wilson M and Jones, Christopher M and Baldwin, Grant T},
  journal={New England Journal of Medicine},
  volume={374},
  number={2},
  pages={154--163},
  year={2016},
  publisher={Mass Medical Soc}
}

@article{dasgupta2014observed,
  title={Observed transition from opioid analgesic deaths toward heroin},
  author={Dasgupta, Nabarun and Creppage, Kathleen and Austin, Anna and Ringwalt, Christopher and Sanford, Catherine and Proescholdbell, Scott K},
  journal={Drug and alcohol dependence},
  volume={145},
  pages={238--241},
  year={2014},
  publisher={Elsevier}
}

@article{cassidy2014changes,
  title={Changes in prevalence of prescription opioid abuse after introduction of an abuse-deterrent opioid formulation},
  author={Cassidy, Theresa A and DasMahapatra, Pronabesh and Black, Ryan A and Wieman, Matthew S and Butler, Stephen F},
  journal={Pain Medicine},
  volume={15},
  number={3},
  pages={440--451},
  year={2014}
}

@article{havens2014impact,
  title={The impact of a reformulation of extended-release oxycodone designed to deter abuse in a sample of prescription opioid abusers},
  author={Havens, Jennifer R and Leukefeld, Carl G and DeVeaugh-Geiss, Angela M and Coplan, Paul and Chilcoat, Howard D},
  journal={Drug and alcohol dependence},
  volume={139},
  pages={9--17},
  year={2014},
  publisher={Elsevier}
}

@article{o2017trends,
  title={Trends in deaths involving heroin and synthetic opioids excluding methadone, and law enforcement drug product reports, by census region—United States, 2006--2015},
  author={O’Donnell, Julie and Gladden, Matthew and Seth, Puja},
  journal={MMWR. Morbidity and mortality weekly report},
  volume={66},
  number={34},
  pages={897},
  year={2017},
  publisher={Centers for Disease Control and Prevention}
}

@article{pergolizzi2018abuse,
  title={Abuse-deterrent opioids: an update on current approaches and considerations},
  author={Pergolizzi, Joseph and Raffa, Robert  and Taylor Jr, Robert and Vacalis, Steven},
  journal={Current medical research and opinion},
  volume={34},
  number={4},
  pages={711--723},
  year={2018},
  publisher={Taylor \& Francis}
}

@article{carpenter2017economic,
  title={Economic conditions, illicit drug use, and substance use disorders in the United States},
  author={Carpenter, Christopher S and McClellan, Chandler B and Rees, Daniel I},
  journal={Journal of Health Economics},
  volume={52},
  pages={63--73},
  year={2017},
  publisher={Elsevier}
}

@article{swensen2015substance,
  title={Substance-abuse treatment and mortality},
  author={Swensen, Isaac D},
  journal={Journal of Public Economics},
  volume={122},
  pages={13--30},
  year={2015},
  publisher={Elsevier}
}

@article{cicero2014changing,
  title={The changing face of heroin use in the United States: a retrospective analysis of the past 50 years},
  author={Cicero, Theodore J and Ellis, Matthew S and Surratt, Hilary L and Kurtz, Steven P},
  journal={JAMA psychiatry},
  volume={71},
  number={7},
  pages={821--826},
  year={2014},
  publisher={American Medical Association}
}

@article{kennedy2016opioid,
  title={Opioid overdose deaths and Florida’s crackdown on pill mills},
  author={Kennedy-Hendricks, Alene and Richey, Matthew and McGinty, Emma and Stuart, Elizabeth and Barry, Colleen and Webster, Daniel},
  journal={American journal of public health},
  volume={106},
  number={2},
  pages={291--297},
  year={2016},
  publisher={American Public Health Association}
}

@article{meier2007guilty,
  title={In guilty plea, OxyContin maker to pay \$600 million},
  author={Meier, Barry},
  journal={New York Times},
  volume={10},
  year={2007}
}

@article{finkelstein2007aggregate,
  title={The aggregate effects of health insurance: Evidence from the introduction of Medicare},
  author={Finkelstein, Amy},
  journal={The quarterly journal of economics},
  volume={122},
  number={1},
  pages={1--37},
  year={2007},
  publisher={MIT Press}
}

@book{meier2003pain,
  title={Pain killer: A" wonder" drug's trail of addiction and death},
  author={Meier, Barry},
  year={2003},
  publisher={Rodale}
}

@book{macy2018dopesick,
  title={Dopesick: Dealers, doctors, and the drug company that addicted America},
  author={Macy, Beth},
  year={2018},
  publisher={Little, Brown}
}

@book{quinones2015dreamland,
  title={Dreamland: The true tale of America's opiate epidemic},
  author={Quinones, Sam},
  year={2015},
  publisher={Bloomsbury Publishing USA}
}

@article{leung20171980,
  title={A 1980 letter on the risk of opioid addiction},
  author={Leung, Pamela TM and Macdonald, Erin M and Stanbrook, Matthew B and Dhalla, Irfan A and Juurlink, David N},
  journal={New England Journal of Medicine},
  volume={376},
  number={22},
  pages={2194--2195},
  year={2017},
  publisher={Mass Medical Soc}
}

@article{deweerdt2019tracing,
  title={Tracing the US opioid crisis to its roots.},
  author={DeWeerdt, Sarah},
  journal={Nature},
  volume={573},
  number={7773},
  pages={S10--S10},
  year={2019},
  publisher={Nature Publishing Group}
}

@article{adler2018overview,
  title={An overview of abuse-deterrent opioids and recommendations for practical patient care},
  author={Adler, Jeremy A and Mallick-Searle, Theresa},
  journal={Journal of multidisciplinary healthcare},
  volume={11},
  pages={323},
  year={2018},
  publisher={Dove Press}
}
